%% file: main.tex
\documentclass{LMCS}

\usepackage{amsmath,amssymb}
\usepackage{xspace}
\usepackage{stmaryrd}
\usepackage{multirow}
\usepackage{graphicx}
\usepackage{hyperref}

\input{./definitions}

\allowdisplaybreaks

\def\doi{3 (2:7) 2007}
\lmcsheading%
{\doi}
{1--33}
{}
{}
{Oct.~25, 2006}
{Jun.~29, 2007}
{}   

\begin{document}

\title[The Complexity of Model Checking HFL]{The Complexity of Model Checking Higher-Order Fixpoint Logic}

\author[R.~Axelsson]{Roland Axelsson\rsuper a}
\address{{\lsuper a}Department of Computer Science, University of Munich, Germany}
\email{Roland.Axelsson@ifi.lmu.de}

\author[M.~Lange]{Martin Lange\rsuper b}
\address{{\lsuper b}Department of Computer Science, University of Aarhus, Denmark}
\email{Martin.Lange@ifi.lmu.de}

\author[R.~Somla]{Rafa\l{} Somla\rsuper c}
\address{{\lsuper c}IT Department, Uppsala University, Sweden}
\email{Rafal.Somla@it.uu.se}

\titlecomment{}

\begin{abstract}

Higher-Order Fixpoint Logic (\HFL) is a hybrid of the simply typed
$\lambda$-calculus and the modal $\mu$-calculus. This makes it a
highly expressive temporal logic that is capable of expressing various
interesting correctness properties of programs that are not
expressible in the modal $\mu$-calculus.

This paper provides complexity results for its model checking
problem. In particular, we consider those fragments of \HFL that are
built by using only types of bounded order $k$ and arity $m$.  We
establish $k$-fold exponential time completeness for model checking each
such fragment. For the upper bound we use fixpoint elimination to
obtain reachability games that are singly-exponential in the size of
the formula and $k$-fold exponential in the size of the underlying
transition system. These games can be solved in deterministic linear
time. As a simple consequence, we obtain an exponential time upper
bound on the expression complexity of each such fragment.

The lower bound is established by a reduction from the word problem
for alternating $(k-1)$-fold exponential space bounded Turing
Machines. Since there are fixed machines of that type whose word
problems are already hard with respect to $k$-fold exponential time,
we obtain, as a corollary, $k$-fold exponential time completeness for
the data complexity of our fragments of \HFL, provided $m$ exceeds 3.
This also yields a hierarchy result in expressive power.

\end{abstract}

\maketitle

\input{intro}

\input{prelim}
\input{upper}
\input{lower}

\input{concl}

\bibliographystyle{plain}
\bibliography{./literature}

\end{document}

%% file: definitions.tex
\def\follows{\vdash}
\def\Pr{\mathsf{Pr}}
\def\<#1>{\langle#1\rangle}

\def\semb#1{[\![ #1 ]\!]}

\def\fn{\lambda}

\newcommand{\Nat}{\ensuremath{\mathbb{N}}}

\newcommand{\tower}[2]{\ensuremath{\mathbf{2}^{#1}_{#2}}}

\newcommand{\hfl}[1]{\ensuremath{\mathrm{HFL}^{#1}}}
\newcommand{\mucalc}[1]{\ensuremath{\mathcal{L}_{\mu}^{#1}}}

\newcommand{\true}{\ensuremath{\mathtt{t\!t}}}
\newcommand{\false}{\ensuremath{\mathtt{f\!f}}}
\newcommand{\Mubox}[1]{\ensuremath{[ #1 ]}}
\newcommand{\Mudiam}[1]{\ensuremath{\langle #1 \rangle}}

\newcommand{\height}[1]{\ensuremath{h(#1)}}

\newcommand{\order}[1]{\ensuremath{\mathit{ord}(#1)}}
\newcommand{\mar}[1]{\ensuremath{\mathit{mar}(#1)}}
\newcommand{\types}[1]{\ensuremath{\mathit{types}(#1)}}

\newcommand{\States}{\ensuremath{\mathcal{S}}}
\newcommand{\Labelling}{\ensuremath{L}}
\newcommand{\Transsys}{\ensuremath{\mathcal{T}}}

\def\newarrow#1{\mathop{{\hbox{\setbox0=\hbox{$\scriptstyle{#1\quad}$}{$%
\mathrel{\mathop{\setbox1=\hbox to
\wd0{\rightarrowfill}\ht1=3pt\dp1=-2pt\box1}\limits^{#1}}%
$}}}}}

\newcommand{\Transition}[3]{\ensuremath{#1 \newarrow{#2} #3}}

\newcommand{\game}[3]{\ensuremath{\mathcal{G}_{#1}(#2,#3)}}

\newcommand{\env}{\eta}
\newcommand{\semTphiEta}[1]{\ensuremath{\semb{#1}^\Transsys_\env}}
\newcommand{\semTphiNEta}[2]{\ensuremath{\semb{#1}^\Transsys_{#2}}}
\newcommand{\StateTrans}[1]{\ensuremath{\semb{#1}^\Transsys}}

\newcommand{\repr}[2]{\ensuremath{|\!|#1|\!|^{#2}}}

\newcommand{\ELTIME}{\textsc{Elementary}\xspace}
\newcommand{\DTIME}[1]{\ensuremath{\mathrm{DTime(}#1\mathrm{)}}}

\newcommand{\ExpTime}{\textsc{ExpTime}\xspace}
\newcommand{\EXPTIME}[1]{\ensuremath{#1}\ExpTime}

\newcommand{\ASPACE}[1]{\ensuremath{\mathrm{ASpace(}#1\mathrm{)}}}

\newcommand{\AEXPSPACE}[1]{\ensuremath{{#1}\mathrm{AExpSpace}}}

\newcommand{\HFL}{\ensuremath{\mathrm{HFL}}\xspace}

\newcommand{\FLC}{\ensuremath{\mathrm{FLC}}\xspace}    %

\newcommand{\Prop}{\ensuremath{\mathcal{P}}}
\newcommand{\Act}{\ensuremath{\mathcal{A}}}
\newcommand{\Var}{\ensuremath{\mathcal{V}}}

\newcommand{\fl}[1]{\ensuremath{\mathit{FL}(#1)}}

\let\sig\sigma

\newcommand{\pot}[1]{\ensuremath{2^{#1}}}

\renewcommand{\phi}{\varphi}



%% file: intro.tex
\section{Introduction}

Temporal logics are well-established tools for the specification of correctness 
properties and their verification in hard- and software design processes. One 
of the most famous temporal logics is Kozen's modal $\mu$-calculus \mucalc{} 
\cite{Kozen83} which extends multi-modal logic with extremal fixpoint 
quantifiers. \mucalc{} subsumes many other temporal logics like PDL \cite{Fischer79}
as well as CTL${}^*$ \cite{Emerson:1986:SNN}, and with it CTL \cite{EmersonHalpern85} and LTL 
\cite{Pnueli:1977}. It also has connections to other formalisms like 
description logics for example.

\mucalc{} is equi-expressive to the bisimulation-invariant fragment of Monadic 
Second Order Logic over trees or graphs \cite{focs91*368,CONCUR::JaninW1996}.
Hence, properties expressed by formulas of the modal $\mu$-calculus are only 
regular. There are, however, many interesting correctness properties of 
programs that are not regular. Examples include \emph{uniform inevitability} 
\cite{Emerson:1987:UIT} which states that a certain event occurs globally at 
the same time in all possible runs of the system; counting properties like ``at 
any point in a run of a protocol there have never been more \emph{send}- than 
\emph{receive}-actions''; formulas saying that an unbounded number of data does 
not lose its order during a transmission process; or properties making structural 
assertions about their models like being bisimilar to a linear time model.

When program verification was introduced to computer science, programs as well 
as their correctness properties were mainly specified in temporal logics. 
Hence, verification meant to check formulas of the form $\varphi \to \psi$ for 
validity, or equally formulas of the form $\varphi \wedge \psi$ for 
satisfiability. An intrinsic problem for this approach and non-regular 
properties is undecidability. Note that the intersection problem for 
context-free languages is already undecidable \cite{barhillel-etal:1961a}.

One of the earliest attempts at verifying non-regular properties of programs 
was \emph{Non-Regular PDL} \cite{JCSS::HarelPS1983} which enriches ordinary PDL 
by context-free programs. Non-Regular PDL is highly undecidable, hence, the 
logic did not receive much attention for program verification purposes. Its 
model checking problem, however, remains decidable on finite transition systems
-- it is even in P \cite{lange-mcpdl}.

Another example is  
\emph{Fixpoint Logic with Chop}, FLC, \cite{Mueller-Olm:1999:MFL} which extends 
\mucalc{} with a sequential composition operator. It is capable of expressing 
many non-regular -- and even non-context-free -- properties, and its model 
checking problem on finite transition systems is decidable in deterministic 
exponential time \cite{ls-fossacs02}. It also properly subsumes 
Non-Regular PDL \cite{langesomla-ipl06}. 

In order to achieve non-regular effects in FLC, the original \mucalc{} 
semantics is lifted to a function from sets of states to sets of states. This 
idea has been followed consequently in the introduction of \emph{Higher-Order 
Fixpoint Logic}, \HFL, \cite{viswanathan:hfl} which incorporates a simply typed 
$\lambda$-calculus into the modal $\mu$-calculus. This gives it even more 
expressive power than FLC. \HFL is, for example, capable of expressing 
\emph{assume-guarantee-properties}. Still, \HFL's model checking problem on 
finite transition systems remains decidable. This has been stated in its 
introductory work \cite{viswanathan:hfl}. It is also known that model checking
\HFL is non-elementary with the following complexity bounds \cite{langesomla-mfcs05}.
\begin{itemize}
\item When restricted to function types of order $k$, the model checking problem 
      for this fragment is hard for deterministic $(k-3)$-fold exponential space and included in 
      determinisitic $(k+1)$-fold exponential time. It is not made explicit, though, that the
      arity of types needs to be fixed for that.
\item The model checking problem is non-elementary on fixed (and very small)
      structures already. However, unbounded type orders are needed for this 
      result.
\end{itemize}
Our aim is to close this apparent gap and to provide an analysis of the model checking
problem for \HFL and, thus, the problem of automatically verifying non-regular properties
on finite transition systems.

We start in Sect.~\ref{sec:prel} by 
recalling the logic and giving a few examples of \HFL-expressible properties. 
Sect.~\ref{sec:upper} contains a reduction from \HFL's model checking
problem to the problem of solving (rather large) reachability games. This improves 
the upper bound mentioned above: these games can be solved in $k$-fold exponential 
time when type orders are bounded by $k$ and arities are fixed.

Sect.~\ref{sec:lower} presents a reduction from the word problem for alternating
space-bounded Turing Machines to \HFL's model checking problem. This improves
on the lower bounds mentioned above in two ways. For the fragment
of type orders restricted to $k$ we can match the new upper bound and establish
completeness for the class of $k$-fold deterministic exponential time. A slight
modification produces formulas that are independent of the input word to the
Turing Machine. Hence, we get a result on the data complexity of \HFL as a simple
corollary. This, in turn, yields a hierarchy result on expressive power within
\HFL.

A non-elementary lower complexity bound on the problem of a logic that incorporates
the simply typed $\lambda$-calculus is of course reminiscent of Statman's result
which states that the normalisation problem in the simply typed $\lambda$-calculus 
is non-elementary \cite{statman79tcsb}. But this is rather related to the equivalence
problem for \HFL which is known to be highly undecidable 
\cite{JCSS::HarelPS1983,langesomla-ipl06,viswanathan:hfl}. Since \HFL is a branching
time logic there is probably no simple reduction from the equivalence problem to the 
model checking problem. Hence, the lower bounds presented here do not necessarily 
follow from Statman's result. 

Furthermore, Statman's result is of course irrelevant for the upper bounds presented
here. There is some work on upper bounds for the number of $\beta$-reduction steps in
the simply typed $\lambda$-calculus, c.f.\ \cite{Schwichtenberg91}. However, this is not good enough to obtain the upper bounds we are after, c.f.\ 
Sect.~\ref{sec:upper}. It also does not deal with the propositional, modal and 
fixpoint parts of \HFL formulas.


%% file: prelim.tex
\section{Preliminaries}
\label{sec:prel}

\subsection{The Syntax of Formulas}

\begin{defi}
Let $\Prop = \{ p, q, \ldots\}$ be a set of atomic propositions, $\Act =\{ a, 
b, \ldots\}$ be a finite set of action names, and $\Var=\{X, Y, \ldots\}$ 
a set of variables. For simplicity, we fix $\Prop$, $\Act$, and $\Var$ for 
the rest of the paper.

A $v\in\{-,+,0\}$ is called a variance. The set of \HFL types is the smallest 
set containing the atomic type $\Pr$ and being closed under function typing 
with variances, i.e.\ if $\sig$ and $\tau$ are \HFL types and $v$ is a 
variance, then $\sig^v\to \tau$ is an \HFL type. 

Formulas of \HFL are given by the following grammar:
\begin{displaymath}
\varphi \enspace ::= \enspace q \mid X \mid \neg\varphi \mid
\phi\lor\phi \mid \Mudiam{a}\varphi \mid \varphi\ \varphi \mid 
\lambda(X^v:\tau).\varphi \mid \mu(X:\tau).\phi
\end{displaymath}
where $q \in \Prop$, $X \in \Var$, $a \in \Act$, $v$ is a variance and $\tau$ is an \HFL type.

An \HFL formula $\varphi$ is called \emph{fixpoint-free} if it does not contain 
any subformula of the form $\mu X.\psi$.
\end{defi}

Throughout this paper we will adopt the convention given by the syntax of \HFL and
write function application in the style $f\; x$ rather than $f(x)$.

We use the following standard abbreviations:
\begin{displaymath}
\begin{array}{rclrcl}
\true & := & q \vee \neg q \enspace \mbox{for some } q \in \Prop \qquad &
\false  &:= & \neg\true \\
\varphi \wedge \psi  &:= & \neg(\neg\varphi \vee \neg\psi) &
\varphi \to \psi  &:= & \neg\varphi \vee \psi \\
\varphi \leftrightarrow \psi &:=& (\varphi \to \psi) \wedge (\psi \to \varphi) &
\nu X.\varphi  &:= & \neg \mu X.\neg\varphi[\neg X / X] \\
\Mubox{a}\psi  &:= & \neg\Mudiam{a}\neg\psi &
\Mudiam{-}\varphi  &:= & \bigvee_{a \in \Act} \Mudiam{a}\varphi \\
\Mubox{-}\varphi  &:= & \bigwedge_{a \in \Act} \Mubox{a}\varphi & & & \\
\end{array}
\end{displaymath}
where $\varphi[\psi / X]$ denotes the formula that results from $\varphi$ by
replacing simultaneously every occurrence of $X$ by $\psi$.

\begin{figure}[tp]
\framebox[\textwidth]{
\parbox{\textwidth}{
\vspace*{2mm}
\begin{displaymath}
   \begin{array}{c}
   \hspace*{1cm} \\ \hline
    \Gamma\follows q\colon\Pr
   \end{array}
   \hspace*{1.6cm}
   \begin{array}{c}
     v\in\{0,+\} \\\hline
     \Gamma, X^v\colon\tau \follows X\colon\tau
   \end{array}
  \hspace*{1.6cm}
   \begin{array}{c}
     \Gamma^- \follows \phi\colon\tau
     \\\hline
     \Gamma \follows \neg\phi\colon\tau
   \end{array} 
\end{displaymath}
\vspace{2mm}
\begin{displaymath}
   \begin{array}{c}
     \Gamma\follows\phi\colon\Pr \quad \Gamma\follows\psi\colon\Pr 
     \\\hline
     \Gamma\follows\phi\lor\psi\colon\Pr
   \end{array}
   \hspace*{1cm}
  %
   \begin{array}{c}
     \Gamma \follows \phi\colon\Pr
     \\\hline
     \Gamma \follows \Mudiam{a}\phi\colon\Pr
   \end{array}
   \hspace*{1cm}
   \begin{array}{c}
     \Gamma, X^v\colon\sig \follows \phi\colon\tau
     \\\hline
     \Gamma \follows \fn(X^v\colon\sig).\phi : (\sig^v\to\tau)
   \end{array}
\end{displaymath}
\vspace{2mm}
\begin{displaymath}
   \begin{array}{c}
    \Gamma\follows\phi:(\sig^+\to\tau) \quad \Gamma\follows \psi\colon\sig 
    \\\hline
    \Gamma\follows (\phi\;\psi) : \tau
   \end{array}%
   \hspace*{2cm}
   \begin{array}{c}
    \Gamma\follows\phi:(\sig^-\to\tau) \quad \Gamma^-\follows \psi\colon\sig 
    \\\hline
    \Gamma\follows (\phi\;\psi) : \tau
   \end{array}%
\end{displaymath}
\vspace{2mm}
\begin{displaymath}
   \begin{array}{c}
    \Gamma\follows\phi:(\sig^0\to\tau) \quad \Gamma\follows \psi\colon\sig \quad  \Gamma^-\follows \psi\colon\sig
    \\\hline
    \Gamma\follows (\phi\;\psi) : \tau
   \end{array}
   \hspace*{2cm}
   \begin{array}{c}
     \Gamma, X^+\colon\tau \follows \phi\colon\tau
     \\\hline
     \Gamma \follows \mu(X\colon\tau).\phi : \tau
   \end{array}
\end{displaymath}
\vspace*{2mm}
}}

\caption{Type inference rules for HFL.}
\label{type-rules}
\end{figure}

\begin{defi}
A sequence $\Gamma$ of the form $X^{v_1}_1\colon \tau_1, \dots, X^{v_n}_n\colon 
\tau_n$ where $X_i$ are variables, $\tau_i$ are types and $v_i$ are variances 
is called a \emph{context} (we assume all $X_i$ are distinct). An \HFL formula 
$\phi$ has type $\tau$ in context $\Gamma$ if the statement 
$\Gamma\follows\phi\colon\tau$ can be inferred using the rules of 
Fig.~\ref{type-rules}. We say that $\phi$ is \emph{well-formed} if 
$\Gamma\follows\phi\colon\tau$ for some $\Gamma$ and $\tau$.

For a variance $v$, we define its complement $v^-$ as $+$ if $v=-$, as $-$ if 
$v=+$, and $0$ otherwise. For a context $\Gamma = X^{v_1}_1\colon \tau_1, 
\dots, X^{v_n}_n\colon \tau_n$, the complement $\Gamma^-$ is defined as 
$X^{v_1^-}_1\colon \tau_1, \dots, X^{v_n^-}_n\colon \tau_n$.
\end{defi}

\begin{defi}
The Fischer-Ladner closure of an \HFL formula $\varphi_0$ is the least set 
$\fl{\varphi_0}$ that contains $\varphi_0$ and satisfies the following.
\begin{itemize}
\item If $\psi_1 \vee \psi_2 \in \fl{\varphi_0}$ then 
      $\{\psi_1,\psi_2\} \subseteq \fl{\varphi_0}$.
\item If $\neg(\psi_1 \vee \psi_2) \in \fl{\varphi_0}$ then 
      $\{\neg\psi_1,\neg\psi_2\} \subseteq \fl{\varphi_0}$.
\item If $\Mudiam{a}\psi \in \fl{\varphi_0}$ then $\psi \in \fl{\varphi_0}$.
\item If $\neg\Mudiam{a}\psi \in \fl{\varphi_0}$ then $\neg\psi \in \fl{\varphi_0}$.
\item If $\varphi\;\psi \in \fl{\varphi_0}$ then $\{\varphi,\psi,\neg\psi\} \subseteq \fl{\varphi_0}$.
\item If $\neg(\varphi\;\psi) \in \fl{\varphi_0}$ then $\{\neg\varphi,\psi,\neg\psi\} \subseteq \fl{\varphi_0}$.
\item If $\lambda X.\psi \in \fl{\varphi_0}$ then $\psi \in \fl{\varphi_0}$.
\item If $\neg(\lambda X.\psi) \in \fl{\varphi_0}$ then $\neg\psi \in \fl{\varphi_0}$.
\item If $\mu X.\psi \in \fl{\varphi_0}$ then $\psi \in \fl{\varphi_0}$.
\item If $\neg(\mu X.\psi) \in \fl{\varphi_0}$ then $\neg\psi[\neg X/ X] \in \fl{\varphi_0}$.
\item If $\neg\neg \psi \in \fl{\varphi_0}$ then $\psi \in \fl{\varphi_0}$.
\item If $\neg X \in \fl{\varphi_0}$ then $X \in \fl{\varphi_0}$.
\item If $\neg q \in \fl{\varphi_0}$ then $q \in \fl{\varphi_0}$.
\end{itemize}

Note that the size of $\fl{\varphi}$ as a set is at most twice the length of $\varphi$. We
therefore define $|\varphi| := |\fl{\varphi}|$. Another measure for the complexity of a formula
is the number $v(\varphi)$ of distinct $\lambda$-bound variables occurring in $\varphi$. Formally,
let $v(\varphi) := | \{ X \mid \lambda X.\psi \in \fl{\varphi}$ for some $\psi
\} | $.\footnote{Note
that we do not require $\alpha$-equivalent formulas to have exactly the same computational measures.}
\end{defi}






When using least fixpoint quantifiers it is often beneficial to recall the B\'eki\`c principle \cite{Bekic84} 
which states that a simultaneously defined least fixpoint of a monotone function is the same as a parametrised one.
We will use this to allow formulas like
\begin{displaymath}
\varphi \enspace := \enspace \mu X_i.\left(
\begin{array}{lcl}
X_1 & . & \varphi_1(X_1,\ldots,X_n) \\
 &\ \vdots & \\
X_n & . & \varphi_n(X_1,\ldots,X_n)
\end{array}\right)
\end{displaymath}
in the syntax of \HFL. This abbreviates  
\begin{displaymath}
\varphi' \enspace := \enspace 
\mu X_i.\varphi_i(\mu X_1.\varphi_1(X_1,\mu X_2.\varphi_2(X_1,X_2,\ldots,X_i,\ldots),\ldots,X_i,\ldots),\mu X_2\ldots,\ldots,X_i,\ldots)
\end{displaymath}
Note that the size of $\varphi'$ can be exponentially bigger than the size of $\varphi$, and this even holds for the number
of their subformulas. However, it is only exponential in $n$, not in $|\varphi|$: $|\varphi'| = O(|\varphi|\cdot 2^n)$.

\subsection{The Semantics of Types and Formulas}

\begin{defi}
A (labeled) transition system is a structure $\Transsys = 
(\States,\{\Transition{}{a}{} \mid a \in \Act\},\Labelling)$ where $\States$ is a finite non-empty set of 
states, $\Transition{}{a}{}$ is a binary relation on states for each $a \in 
\Act$, and $\Labelling : \States \to \pot{\Prop}$ is a function labeling each state
with the set of propositional constants that are true in it.

The semantics of a type w.r.t.\ a transition system $\Transsys$ is a Boolean
lattice\footnote{In the original definition, the semantics is only said to be a complete lattice
but it is in fact also Boolean. The reason for this is that negation is only allowed on the ground 
type $\Pr$ anyway. The game-based characterisation of \HFL's model checking problem in the following section 
benefits from a symmetric definition w.r.t.\ negation. Hence, we allow negation in the syntax on arbitrary
type levels. But then we have to also use the property of being Boolean of the complete
lattices that form the basis for the definition of the semantics.}, inductively defined on the type as
\[ 
  \StateTrans{\Pr} =  (\pot{\States}, \sqsubseteq_\Pr) \;, \qquad
  \StateTrans{\sigma^v \to \tau} =  
    \big((\StateTrans{\sigma})^v \to \StateTrans{\tau},\sqsubseteq_{\sigma^v \to \tau}\big) \;.
\]
where $\sqsubseteq_\Pr$ is simply the set inclusion order $\subseteq$. For two partial orders 
$\bar{\tau} = (\tau, \sqsubseteq_\tau)$ and $\bar\sigma = (\sigma, \sqsubseteq_\sigma)$, 
$\bar\sigma \to \bar\tau$ denotes the partial order of all monotone functions ordered pointwise. I.e., in this case,
\begin{displaymath}
f \sqsubseteq_{\sigma^v \to \tau} g \quad \mbox{iff} \quad \mbox{for all } x \in \StateTrans{\sigma}: f\; x \sqsubseteq_{\tau} g\; x
\end{displaymath}
Moreover, complements in these lattices are denoted by $\bar{f}$ and defined on higher levels as
$\bar{f}\; x = \overline{f\; x}$. 

A positive variance leaves a partial order unchanged, $\bar\tau^+ = (\tau, \sqsubseteq_\tau)$, 
a negative variance turns it upside-down to make antitone functions look well-behaved,
$\bar\tau^- = (\tau, \sqsupseteq_\tau)$, and a neutral variance flattens it,
$\bar\tau^0 = (\tau, \sqsubseteq_\tau \cap \sqsupseteq_\tau)$. This is not a complete lattice
anymore which does not matter since variances only occur on the left of a typing arrow. Note that 
the space of monotone functions from a partial order to a Boolean lattice with pointwise ordering forms a 
Boolean lattice again.
\end{defi}

\begin{figure}[t]
\noindent\framebox[\textwidth]{
\parbox{\textwidth}{
\begin{center}
\begin{math}
\begin{array}{rcl}
\semTphiEta{\Gamma \vdash q  : \Pr} & = & 
  \{ s \in \States \mid q \in  \Labelling(s)\} \\[1ex]
\semTphiEta{\Gamma \vdash X : \tau} & = & \env(X) \\[1ex]
%
%
\semTphiEta{\Gamma \vdash \neg \phi : \Pr} & = & \States \setminus 
  \semTphiEta{\Gamma^- \vdash \phi : \Pr} \\[1ex]
\semTphiEta{\Gamma \vdash \neg \phi : \sigma^v \to \tau} & = & 
  f \in \StateTrans{\sigma^v \to \tau} \mbox{ s.t. } \bar{f} = \semTphiEta{\Gamma^- \vdash \phi: \sigma^v \to \tau} \\[1ex]
\semTphiEta{\Gamma \vdash \phi \vee \psi : \Pr} & = & 
  \semTphiEta{\Gamma \vdash \phi : \Pr} \cup
  \semTphiEta{\Gamma \vdash \psi : \Pr} \\[1ex]
\semTphiEta{\Gamma \vdash \Mudiam{a}\phi : \Pr} & = & 
  \{ s \in \States \mid \Transition{s}{a}{t} \mbox{ for some } 
     t \in \semTphiEta{\Gamma \vdash \phi : \Pr}\} \\[1ex]
\semTphiEta{\Gamma \vdash \lambda(X^v : \sigma). \phi : \sigma^v \to \tau} & = & 
  f \in \StateTrans{\sigma^v \to \tau} \mbox{ s.t. } \forall x \in \StateTrans{\sigma}\\[1mm]
  & & \quad f\; x = \semTphiNEta{\Gamma, X^v : \sigma  \vdash \phi : \tau}{\env[X \mapsto x]} \\[1ex]
\semTphiEta{\Gamma \vdash \phi\; \psi : \tau} & = & 
  \semTphiEta{\Gamma \vdash \phi : \sigma^v \to \tau}\; \semTphiEta{\Gamma' \vdash \psi : \sigma} \\[1ex]
\semTphiEta{\Gamma \vdash \mu(X : \tau) \phi : \tau } & = & 
  \bigsqcap \{x \in \StateTrans{\tau} \mid \semTphiNEta{\Gamma, X^+ : \tau  \vdash \phi : \tau}{\env[X \mapsto x]} \sqsubseteq_{\tau} x \}\\
\end{array}
\end{math}
\end{center}
}}
\caption{Semantics of HFL\label{fig:semantics}}
\end{figure}

\begin{defi}
An \emph{environment} $\env$ is a possibly partial map on the variable set $\Var$. For 
a context $\Gamma = X^{v_1}_1 : \tau_1, \dots, X^{v_n}_n: \tau_n$, we say that 
$\env$ respects $\Gamma$, denoted by $\env \models \Gamma$, if $\env(X_i) \in 
\StateTrans{\tau_i}$ for $i \in \{1, \dots, n\}$. We write $\env[X \mapsto f]$ 
for the environment that maps $X$ to $f$ and otherwise agrees with $\env$. If 
$\env \models \Gamma$ and $f \in \StateTrans{\tau}$ then $\env[X \mapsto f] 
\models \Gamma, X : \tau$, where $X$ is a variable that does not appear in 
$\Gamma$.

For any well-typed term $\Gamma \follows \phi : \tau$ and environment
$\env \models \Gamma$, Fig.~\ref{fig:semantics} defines the
semantics of $\phi$ inductively to be an element of
$\StateTrans{\tau}$.
In the clause for function application $(\phi\; \psi)$ the
context $\Gamma'$ is $\Gamma$ if $v \in \{ +, 0\}$, and is $\Gamma^-$ if $v = -$.

The model checking problem for $\HFL$ is the following: Given an $\HFL$ 
sentence $\phi:\Pr$, a transition system $\Transsys$ and one of its states $s$, decide 
whether or not $s \in \semb{\phi}^\Transsys$.
\end{defi}

In the following we will identify a type $\tau$ and its underlying complete 
lattice $\StateTrans{\tau}$ induced by a transition system $\Transsys$ with state set $\States$. In 
order to simplify notation we fix $\Transsys$ for the remainder of this section. We will also simply
write $|\tau|$ instead of $|\StateTrans{\tau}|$ for the size of the lattice induced by $\tau$.

\begin{defi}
We consider fragments of formulas that can be built using restricted types only. Note that because
of right-associativity of the function arrow, every \HFL type is isomorphic to a 
$\tau = \tau_1 \to \ldots \to \tau_m \to \Pr$ where $m \in \Nat$. Clearly, for $m=0$ we simply
have $\tau = \Pr$. We stratify types w.r.t.\ their \emph{order}, i.e.\ the degree of using proper functions
as arguments to other functions, as well as \emph{maximal arity}, i.e.\ the number of arguments a function
has. Order can be seen as depth, and maximal arity as the width of a type. Both are defined
recursively as follows.
\begin{align*}
\order{\tau_1 \to \ldots \to \tau_m \to \Pr} \enspace &:= \enspace \max \{ 1 + \order{\tau_i} \mid i = 1,\ldots,m \} \\
\mar{\tau_1 \to \ldots \to \tau_m \to \Pr} \enspace &:= \enspace \max ( \{ m \} \cup \{ \mar{\tau_i} \mid i = 1,\ldots,m \} )
\end{align*}
where we assume $\max \emptyset = 0$. Now let, for $k \ge 1$ and $m \ge 1$,
\begin{align*}
\hfl{k,m} \enspace &:= \enspace \{ \ \varphi \in \HFL \mid \ \vdash \varphi:\Pr \mbox{ using types } \tau \mbox{ with } 
\order{\tau} \le k \mbox{ and } \mar{\tau} \le m \mbox{ only } \} \\
\hfl{k} \enspace &:= \enspace \bigcup\limits_{m \in \Nat} \hfl{k,m}
\end{align*}
Note that no formula can have maximal type order $k > 0$ but maximal type arity $m = 0$. The combination $k=0$
and $m > 0$ is also impossible. Hence, we define
\begin{displaymath}
\hfl{0} = \{ \ \varphi \in \HFL \mid \ \vdash \varphi:\Pr \mbox{ using types } \tau \mbox{ with } 
\order{\tau} = 0 \mbox{ only } \}
\end{displaymath}
We extend these measures to formulas in a straightforward way: $\order{\varphi} = k$ and $\mar{\varphi} = m$ iff 
$k$ and $m$ are the least $k'$ and $m'$ s.t.\ $\varphi$ can be shown to have some type using types $\tau$ with
$\order{\tau} \le k'$ and $\mar{\tau} \le m'$ only.
\end{defi}

\begin{prop}
\hfl{0} = \mucalc{}.
\end{prop}

\proof
An \hfl{0} formula cannot have any subformula of the form $\lambda X.\psi$ or $\varphi\; \psi$. But deleting these
two clauses from the definition of \HFL's syntax yields exactly the syntax of \mucalc{}. It is not hard to see that
this is faithful, i.e.\ the semantics of this logic regarded as a fragment of \HFL is the same as the semantics
of \mucalc{}.
\qed

\subsection{Examples of Properties Expressible in HFL}

\begin{exa}
\label{example1}
\HFL can express the non-regular (but context-free) property ``on any path the 
number of $\mathit{out}$'s seen at any time never exceeds the number of $\mathit{in}$'s seen so 
far.'' Let
\begin{align*}
\varphi \enspace := \enspace \mu (X:\Pr \to \Pr).(\lambda (Z:\Pr).\Mudiam{out} Z \vee \Mudiam{in}(X\;(X \; Z)))\true
\end{align*}
This formula is best understood by comparing it to the CFG $X \to \mathit{out}\mid \mathit{in}\,X\,X$. It generates 
the language $L$ of all words $w \in \{\mathit{in},\mathit{out}\}^*\{\mathit{out}\}$ s.t.\ 
$|w|_{\mathit{in}} = |w|_{\mathit{out}}$ and for all prefixes $v$ of $w$ we have: 
$|v|_{\mathit{in}} \ge |v|_{\mathit{out}}$. This language contains exactly those prefixes of buffer runs that are 
violating due to a buffer underflow. Then $\Transsys,s \models \varphi$ iff there is a finite path through 
$\Transsys$ starting in $s$ that
is labeled with a word in $L$, and $\neg\varphi$ consequently describes the property mentioned above.
\end{exa}

\begin{exa}
Another property that is easily seen not to be expressible by a finite tree automaton
and, hence, not by a formula of the modal $\mu$-calculus either is \emph{bisimilarity to a word}.
Note that a transition system $\Transsys$ with starting state $s$ is not bisimilar to
a linear word model iff there are two distinct actions $a$ and $b$ s.t.\ there are two
(not necessarily distinct) states $t_1$ and $t_2$ at the same distance from $s$ s.t.\ 
$\Transition{t_1}{a}{t_1'}$ and $\Transition{t_2}{b}{t_2'}$ for some $t_1',t_2'$. This
is expressed by the \HFL formula
\begin{displaymath}
\neg \Big(\bigvee_{a \ne b} \big(\mu (F:\Pr\to\Pr\to\Pr). \lambda (X:\Pr).\lambda (Y:\Pr).(X \wedge Y) \vee (F\; \Mudiam{-} X\; \Mudiam{-} Y)\big)\; 
\Mudiam{a}\true\; \Mudiam{b}\true \ \Big)
\end{displaymath}
This formula is best understood by
regarding the least fixpoint definition $F$ as a functional program. It takes two
arguments $X$ and $Y$ and checks whether both hold now or calls itself recursively with the
arguments being checked in two (possibly different) successors of the state that it
is evaluated in.

Note that here, bisimulation does not consider the labels of states but only
the actions along transitions. It is not hard to change the formula accordingly to 
incorporate state labels as well.
\end{exa}

\begin{exa}
Let $\tower{n}{0} := n$ and $\tower{n}{m+1} := 2^{\tower{n}{m}}$. For any $m 
\in \Nat$, there is a short \HFL formula $\varphi_m$ expressing the fact that there 
is a maximal path of length $\tower{1}{m}$ (number of states on this path) 
through a transition system. It can be constructed using a typed version of the 
Church numeral $2$. Let $\tau_0 = \Pr$ and $\tau_{i+1} = \tau_i\to \tau_i$. For $i\geq 1$ 
define $\psi_i$ of type $\tau_{i+1}$ as 
$\lambda(F:\tau_i).\lambda(X:\tau_{i-1}).F\,(F\,X)$. Then
\[
   \phi_m := \psi_m\;\psi_{m-1}\;\dots\;\psi_1\;\big(\lambda{(X:\Pr)}.\Mudiam{-} X\big)\;
    \Mubox{-} \false \;.
\]
Note that for any $m \in \Nat$, $\varphi_m$ is of size linear in $m$. This 
indicates that \HFL is able to express computations of Turing Machines of 
arbitrary elementary complexity. Sect.~\ref{sec:lower} will show that this is indeed the 
case.
\end{exa}

\subsection{Complexity Classes and Alternating Turing Machines}

We will assume familiarity with the concept of a deterministic Turing Machine but quickly recall the
less known model of an alternating Turing Machine.

Let $\DTIME{f(n)}$ be the class of languages that can be recognised by a deterministic Turing 
Machine in at most $f(n)$ many steps on any input of length $n$. The $k$-th level of the 
exponential time hierarchy for $k \in \Nat$ is 
\begin{displaymath}
\EXPTIME{k} \enspace := \bigcup\limits_{p \mbox{\scriptsize\ polynomial}}\hspace*{-3mm}\DTIME{\tower{p(n)}{k}}
\end{displaymath}
Then $\ELTIME := \bigcup_{k \in \Nat} \EXPTIME{k}$ is the class of problems that can be solved in
elementary time. Note that \ELTIME does not have complete problems because their existence would
lead to a collapse of the hierarchy which is not the case.

\begin{defi}
An \emph{alternating Turing Machine} is a tuple $\mathcal{M} = (Q,\Sigma,\Gamma,q_0,\delta,q_{\mathit{acc}},q_{\mathit{rej}})$ s.t.\ 
its state set $Q$ is partitioned into \emph{existential} states $Q_\exists$, \emph{universal} states $Q_\forall$ and the halting states
$\{q_{\mathit{acc}},q_{\mathit{rej}}\}$. The starting state $q_0$ is either existential or universal. The input alphabet 
$\Sigma$ is a subset of the tape alphabet $\Gamma$ containing a special blank symbol $\Box$. The transition relation is of type 
$Q \times \Gamma \times Q \times \Gamma \times \{-1,0,+1\}$.

$\mathcal{M}$ is called $f(n)$-space bounded for some function $f(n)$ if it never uses more than $f(n)$ many tape cells in a 
computation on a word of length $n$. A configuration of such an $\mathcal{M}$ is a triple 
$C \in Q \times \{0,\ldots,f(n)-1\} \times \Gamma^{f(n)}$
representing the current state, the position of the tape head and the content of the tape. The starting configuration is
$C_0 := (q_0,0,w\Box\ldots\Box)$. A configuration $(q,i,v)$ is called 
\begin{itemize}
\item \emph{existential} if $q \in Q_\exists$,
\item \emph{universal} if $q \in Q_\forall$,
\item \emph{accepting} if $q = q_{\mathit{acc}}$,
\item \emph{rejecting} if $q = q_{\mathit{rej}}$.
\end{itemize}
The computation of $\mathcal{M}$ on $w$ is a tree whose root is $C_0$ s.t.\ an existential configuration has exactly
one successor configuration in the tree, all possible successor configurations of a universal configurations are present in
the tree, and leaves are exactly those configurations that are accepting or rejecting. The successor relation on configurations
is the usual one built on the transition relation $\delta$.

W.l.o.g.\ we can assume that every path of any computation tree of $\mathcal{M}$ on any $w$ will eventually reach an accepting
or rejecting configuration. I.e.\ computation trees are always finite. This can be achieved for example by running an additional
clock which causes a transition to the rejecting state when a configuration has been reached repeatedly.

A computation is called \emph{accepting} if all of its leaves are
accepting. The machine $\mathcal{M}$ accepts the word $w \in L(\mathcal{M})$, if there is an accepting computation tree of $\mathcal{M}$ on $w$.
\end{defi}

Let $\ASPACE{f(n)}$ be the class of languages that can be recognised by an $f(n)$-space bounded alternating Turing Machine.
\begin{displaymath}
\AEXPSPACE{k} \enspace := \enspace \bigcup\limits_{p \mbox{\scriptsize\ polynomial}}\hspace*{-3mm}\ASPACE{\tower{p(n)}{k}}
\end{displaymath}
There is a direct correspondence between the levels of the elementary time hierarchy and classes defined by alternating space-bounded
Turing Machines. For all $k \ge 1$ we have \EXPTIME{k} = \AEXPSPACE{(k-1)} \cite{Chandra:1981:A}. We will make use of a related
result.

\begin{thm}[\cite{Chandra:1981:A}]
\label{thm:alternation}
For every $k \ge 1$ there is a polynomial $p(n)$ and some alternating $\tower{p(n)}{k-1}$-space bounded Turing Machine $\mathcal{A}_k$ 
s.t.\ $L(\mathcal{A}_k)$ over a binary alphabet is \EXPTIME{k}-hard.
\end{thm}

Finally, we need to introduce the class UP -- a subclass of NP. UP consists of all problems that are solvable by a non-deterministic 
polynomial time bounded Turing Machine with at most one accepting computation. As usual, co-UP denotes the complement of UP.
Later we will briefly mention the class UP$\cap$co-UP. Note that UP$\cap$co-UP does not have complete problems either.


%% file: upper.tex
\section{The Upper Bound}
\label{sec:upper}

We will take two steps in order to obtain a \EXPTIME{k} upper bound on the model checking problem for
\hfl{k,m} for every $m \in \Nat$. First we eliminate fixpoint constructs from the formula w.r.t.\ the
underlying transition system. This results in a possibly $k$-fold exponentially larger modal formula 
with $\lambda$-abstractions and function applications. We then reduce the model checking problem for 
such formulas to the reachability game problem in graphs of roughly the same size.

The combination of the elimination step and the reduction step is necessary to achieve the \EXPTIME{k}
upper bound. It would be easy to eliminate the $\lambda$-calculus part from a fixpoint-free formula
using $\beta$-reduction. However, the best known upper bounds on the number of reduction steps in the
simply typed $\lambda$-calculus are approximately of the order $\tower{O(n)}{k+1}$ \cite{Schwichtenberg91} 
which would only yield a $\EXPTIME{(2k+1)}$ upper bound. 

The reason for avoiding the additional $k+1$ exponents is that $\beta$-reduction is a purely syntactical
procedure. We incorporate semantics into these reachability games by evaluating $\lambda$-bound variables
to real functions of finite domain and co-domain rather than unwinding the entire syntactical definition
of that function as a program in the simply typed $\lambda$-calculus. Note that such a function can be
represented by more than one $\lambda$-term. Whereas equivalence of fixpoint-free \HFL formulas is difficult
to decide -- in fact, it is undecidable in general and might require $\beta$-reduction on a fixed transition
system -- it is easy to decide for unique semantical representations of these functions. 

On the other hand, extending the reachability games to games that capture full \HFL formulas including 
fixpoint quantifiers and variables is not easy either, see the example after the definition of the games below.

\subsection{Fixpoint Elimination}

\begin{lem}
\label{lem:typesize} 
For all \HFL types $\tau$ and all transition systems $\Transsys$ with 
$n$ states we have: $|\tau| \le \tower{n\cdot(\mar{\tau}+\order{\tau})^{\order{\tau}}}{\order{\tau}+1}$.
\end{lem}

\proof
We prove this by induction on the structure of $\tau$. Note that there are $2^n$ many different
elements of type $\Pr$, and $\order{\Pr} = 0 = \mar{\Pr}$ which immediately yields the
base case.

For the other cases let $\tau = \tau_1 \to \ldots \to \tau_m \to \Pr$. With uncurrying
it is easy to regard this as a function that takes $m$ arguments of corresponding type
and delivers something of type $\Pr$. Then we have
\begin{align*}
|\tau| \enspace &= \enspace |\tau_1 \to \ldots \to \tau_m \to \Pr| \enspace = \enspace
|\Pr|^{\prod_{i=1}^m |\tau_i|} \enspace = \enspace 2^{n \cdot \prod_{i=1}^m |\tau_i|}\hspace*{-4cm} \\
&= \enspace 2^{n \cdot \prod\limits_{i=1}^m \tower{n\cdot(\mar{\tau_i}+\order{\tau_i})^{\order{\tau_i}}}{\order{\tau_i}+1}} 
  &\mbox{by the hypothesis} \\
&\le \enspace 2^{n \cdot \prod\limits_{i=1}^m \tower{n\cdot(\mar{\tau_i}+\order{\tau}-1)^{\order{\tau}-1}}{\order{\tau}}}
  &\mbox{because of } \order{\tau_i} \le \order{\tau}-1 \\
&\le \enspace 2^{n \cdot \prod\limits_{i=1}^m \tower{n\cdot(\mar{\tau}+\order{\tau}-1)^{\order{\tau}-1}}{\order{\tau}}}
  &\mbox{because of } \mar{\tau_i} \le \mar{\tau} \\
&= \enspace 2^{n \cdot (\tower{n\cdot(\mar{\tau}+\order{\tau}-1)^{\order{\tau}-1}}{\order{\tau}})^m} \\
&= \enspace 2^{n \cdot 2^{m\cdot \tower{n\cdot(\mar{\tau}+\order{\tau}-1)^{\order{\tau}-1}}{\order{\tau}-1}}}
  &\mbox{because of } \order{\tau} \ge 1 \\
&\le \enspace 2^{n \cdot 2^{\tower{n\cdot m\cdot(\mar{\tau}+\order{\tau}-1)^{\order{\tau}-1}}{\order{\tau}-1}}} \\
&\le \enspace 2^{n \cdot 2^{\tower{n\cdot(\mar{\tau}+\order{\tau}-1)^{\order{\tau}}}{\order{\tau}-1}}}
  &\mbox{because of } m \le \mar{\tau}, \order{\tau} \ge 1 \\
&= \enspace 2^{n \cdot \tower{n\cdot(\mar{\tau}+\order{\tau}-1)^{\order{\tau}}}{\order{\tau}}} \\
&\le \enspace 2^{\tower{n\cdot(\mar{\tau}+\order{\tau}-1)^{\order{\tau}}+ \log n}{\order{\tau}}}
  &\mbox{because of } \order{\tau} \ge 1 \\
&\le \enspace 2^{\tower{n\cdot((\mar{\tau}+\order{\tau}-1)^{\order{\tau}}+1)}{\order{\tau}}} \\
&\le \enspace 2^{\tower{n\cdot(\mar{\tau}+\order{\tau})^{\order{\tau}}}{\order{\tau}}}
  &\mbox{because of } \order{\tau} \ge 1 \\
&= \enspace \tower{n\cdot(\mar{\tau}+\order{\tau})^{\order{\tau}}}{\order{\tau}+1}
\end{align*}
which proves the claim.
\qed

Let $\types{k,m} := \{ \tau \mid \order{\tau} \le k, \mar{\tau} \le m\}$ denote the set of types of
restricted order and maximal arity. As mentioned above we have $|\types{0,0}| = 1$, and $|\types{k,0}| = |\types{0,m}| = 0$
for any $k,m \ge 1$.

\begin{lem}
\label{lem:numberoftypes}
For all $k \ge 1$ and $m \ge 1$ we have $|\types{k,m}| \le m^{k\cdot(m^{k-1})}$.
\end{lem}

\proof
By induction on $k$. First consider the case of $k=1$. All types of order $1$ and maximal arity $m$ are of the form
\begin{displaymath}
\tau \enspace := \enspace \underbrace{\Pr \to \ldots \to \Pr}_{i \mbox{\small\ times}} \to \Pr
\end{displaymath}
with $1 \le i \le m$. Clearly, their number is bounded by $m = m^{1 \cdot (m^0)}$.

Now consider any $k > 1$. Remember that any \HFL type is isomorphic to one of the form
$\tau = \tau_1 \to \ldots \to \tau_i \to \Pr$.
Note that $\order{\tau_j} < \order{\tau}$ for all $j=1,\ldots,i$, and $1 \le i \le m$. Then we have
\begin{align*}
|\types{k,m}| \enspace &= \enspace \sum\limits_{i=1}^{m} |\types{k-1,i}|^i \enspace \le \enspace m \cdot |\types{k-1,m}|^m \enspace
\le \enspace m \cdot (m^{(k-1)m^{k-2}})^m \\
&= \enspace m \cdot m^{(k-1)\cdot m \cdot m^{k-2}} \enspace
= \enspace m \cdot m^{(k-1)\cdot m^{k-1}} \enspace = \enspace m^{(k-1)\cdot m^{k-1}+1} \enspace \le \enspace m^{k\cdot m^{k-1}}
\end{align*}
using the hypothesis for $k-1$.
\qed

\begin{lem}
\label{lem:sizeofalltypes}
For any $k \ge 1$ and  any $m \ge 1$ there are at most $m^{k\cdot(m^{k-1})}\cdot\tower{n\cdot(k+m)^k}{k+1}$ many different
functions $f$ of type $\tau$ with $\order{\tau} \le k$ and $\mar{\tau} \le m$ over a transition system with
$n$ states.
\end{lem}

\proof
Immediately from Lemmas~\ref{lem:typesize} and \ref{lem:numberoftypes}.
\qed

\begin{defi}
Let $\tau$ be any \HFL type. We write $\height{\tau}$ for the \emph{height} of the lattice $\semb{\tau}^{\Transsys}$ over a fixed
transition system $\Transsys$. It is the length of a maximal chain 
\begin{displaymath}
f_0 \enspace \sqsubset_\tau \enspace f_1 \enspace \sqsubset_\tau \enspace f_2 \enspace \sqsubset_\tau \enspace \ldots 
\end{displaymath}
of elements that are properly increasing w.r.t.\ $\sqsubseteq_{\tau}$.
In general this is an ordinal number, but if $|\Transsys| < \infty$ then $\height{\tau} \in \Nat$ for all \HFL types
$\tau$.
\end{defi}

\begin{lem}
\label{lem:typeheight}
For all \HFL types $\tau$ and all transition systems $\Transsys$ with 
$n$ states we have: $\height{\tau} \le (n+1)(\tower{n(\mar{\tau}+\order{\tau}-1)^{\order{\tau}-1}}{\order{\tau}})^{\mar{\tau}}$.
\end{lem}

\proof
First consider the case of $\order{\tau} = 0$. Then $\tau = \Pr$, and it is well-known that the power set lattice of $n$ elements
has height $n+1 = (n+1)\cdot (\tower{n(0+0-1)^{0-1}}{0})^{0}$.

Now suppose $\order{\tau} > 0$ and $\tau = \tau_1 \to \ldots \to \tau_m \to \Pr$ for some $m \ge 1$. Let 
$N := (\tower{n(\mar{\tau}+\order{\tau}-1)^{\order{\tau}-1}}{\order{\tau}})^{\mar{\tau}}$.
According to Lemma~\ref{lem:typesize} there are at most $N$ many different tuples 
$\bar{x} \in \semb{\tau_1}^{\Transsys} \times \ldots \times \semb{\tau_m}^{\Transsys}$ because $\order{\tau_i} \le \order{\tau}-1$
for all $i = 1,\ldots,m$.

Using uncurrying we can regard each $f \in \semb{\tau}^{\Transsys}$ as a function that maps each such $\bar{x}$ to an element of $\Pr$.
Now suppose the claim is wrong. Then there is a chain 
\begin{displaymath}
f_0 \enspace \sqsubset_{\tau} \enspace f_1 \enspace \sqsubset_{\tau} \enspace f_2 \enspace \sqsubset_{\tau} \enspace \ldots \enspace 
\sqsubset_{\tau} \enspace f_{(n+1)N+1}
\end{displaymath}
of functions of type $\tau$. Since each one is strictly greater than the preceeding one there is a sequence $\bar{x}_i$ of tuples s.t.\ 
for $i=0,\ldots,(n+1)N$ we have $f_i\;\bar{x}_i \subsetneq f_{i+1}\;\bar{x}_i$. But remember that there are only $N$ tuples altogether.
By the pidgeon hole principle, one of them must occur at least $(n+1)+1$ many times. Thus, there are 
$0 \le i_1 < \ldots < i_{n+2} \le (n+1)N+1$ s.t.\ $\bar{x}_{i_1} = \bar{x}_{i_2} = \ldots = \bar{x}_{i_{n+2}}$. Let $\bar{x}$ simply
denote this element.

By transitivity of the partial order $\sqsubseteq_{\tau}$ we then have
\begin{displaymath}
f_{i_1}\;\bar{x} \enspace \subsetneq \enspace f_{i_2}\;\bar{x} \enspace \subsetneq \enspace \ldots \enspace \subsetneq \enspace 
f_{i_{n+2}}\;\bar{x}
\end{displaymath}
which contradicts the fact that the height of $\Pr$ is only $n+1$. Hence, the height of $\tau$ must be bounded by $(n+1)N$.
\qed

\begin{defi}
Let $\mu X.\varphi$ be an \HFL formula of type $\tau = \tau_1 \to \ldots \to \tau_m \to \Pr$. We define finite approximants
of this fixpoint formula for all $\alpha \in \Nat$ as follows:
\begin{displaymath}
\mu ^0 X.\varphi \enspace := \enspace \lambda (Z_1: \tau_1^0) \ldots \lambda (Z_m: \tau_m^0).\false \quad, \qquad
\mu^{\alpha+1} X.\varphi \enspace := \enspace \varphi[\mu^\alpha X.\varphi / X]
\end{displaymath}
\end{defi}

The next result is an immediate consequence of the Knaster-Tarski theorem \cite{Tars55}.
 
\begin{lem}
\label{lem:approximants}
Let $\Transsys$ be a transition system with state set $\States$ s.t.\ $|\States| < \infty$. For all \HFL formulas
$\mu (X:\tau).\varphi$ and all environments $\rho$ we have: 
$\semb{\mu (X:\tau).\varphi}^{\Transsys}_{\rho} = \semb{\mu^{\height{\tau}} X.\varphi}^{\Transsys}_{\rho}$.
\end{lem}

The following lemma concerns the size of formulas after fixpoint elimination.

\begin{lem}
\label{lem:fpelimination}
Let $\Transsys$ be a finite transition system with $n$ states, $k,m \ge 1$. For every closed \hfl{k,m} formula $\varphi$ there is a fixpoint-free 
and closed $\varphi' \in$ \hfl{k,m} s.t.\ $\semb{\varphi}^{\Transsys} = \semb{\varphi'}^{\Transsys}$, 
$v(\varphi') \le v(\varphi) + m$, and $|\varphi'| \le |\varphi|\cdot(n+1)^{|\varphi|}\cdot (\tower{n(m+k-1)^{k-1}}{k})^{m\cdot |\varphi|}$.
\end{lem}

\proof
First we prove the existence of such a $\varphi'$ by induction on the number $f$ of different fixpoint subformulas of
$\varphi$. If this is $0$ then simply take $\varphi' := \varphi$.

Suppose $f > 0$. Then $\varphi$ contains at least one subformula $\mu (X:\tau).\psi$ of some type $\tau$ s.t.\ $\psi$ is fixpoint-free. According to
Lemma~\ref{lem:approximants} this $\mu (X:\tau).\psi$ is equivalent to $\mu^{\height{\tau}} X.\psi$ over $\Transsys$. Furthermore,
$\mu^{\height{\tau}} X.\psi$ is fixpoint-free. Let $\varphi'' := \varphi[\mu^{\height{\tau}} X.\psi / \mu (X:\tau).\psi]$. 
Since $\varphi''$ contains less fixpoint subformulas as $\varphi$ we can use the induction hypothesis to obtain a $\varphi'$
that is equivalent to $\varphi''$ over $\Transsys$. Lemma~\ref{lem:approximants} shows that $\varphi''$ is equivalent to 
$\varphi$ over $\Transsys$, hence we have $\semb{\varphi}^{\Transsys} = \semb{\varphi'}^{\Transsys}$. Note that fixpoint elimination
does not create free variables, i.e.\ $\varphi'$ is also closed.

What remains to be shown are the corresponding bounds on the size and number of variables of $\varphi'$. First consider $v(\varphi')$.
The only $\lambda$-bound variables in $\varphi'$ are those that are already $\lambda$-bound in $\varphi$ plus at most
$m$ variables for subformulas of the form $\mu^0 X.\psi = \lambda Z_1\ldots\lambda Z_{m'}.\false$ for some $m' \le m$. Note that
the approximants reuse $\lambda$-bound variables which is semantically sound because the value of an $i$-th approximant as a
function cannot depend on an argument of the $j$-th approximant for some $j \ne i$. The only free variables in each approximant
should be those that are free in $\mu (X:\tau).\psi$ already.

Finally, let $N := (n+1)(\tower{n(m+k-1)^{k-1}}{k})^{m}$. We show by induction on the number $f$ of fixpoint subformulas in $\varphi$ that 
the size of $\varphi'$ is bounded by $(N + 1)^f\cdot|\varphi|$. It should be clear that this implies the claim of the lemma.

This is clearly true for $f=0$. Now let $f > 0$, and first
consider the formula $\varphi'' = \varphi[\mu^{\height{\tau}} X.\psi / \mu (X:\tau).\psi]$ as constructed above. Note that 
$\order{\tau} \le k$, and $\mar{\tau} \le m$, and, according to Lemma~\ref{lem:typeheight}, 
$\height{\tau} \le N$. Therefore, we can estimate the size of the approximant that replaces the fixpoint formula as
$|\mu^{\height{\tau}} X.\psi| \le N \cdot |\psi| + m + 1$. This is because the size of the $0$-th approximant is $m+1$ and the
size of the $(i+1)$-st is always $|\psi|$ plus the size of the $i$-th. Then we have
\begin{displaymath}
|\varphi''| \enspace = \enspace |\varphi| - |\psi| + N\cdot|\psi| + m + 1 \enspace = \enspace |\varphi| + (N-1)\cdot |\psi| + m + 1
\enspace \le \enspace N\cdot |\varphi| + m + 1 \enspace \le \enspace (N+1)\cdot |\varphi|
\end{displaymath}
because the size of a formula $\varphi$ must be strictly greater than the maximal arity of any of its subformulas. Now the
number of fixpoint formulas in $\varphi''$ is $f-1$. By the induction hypothesis we obtain
\begin{displaymath}
|\varphi'| \enspace \le \enspace (N+1)^{f-1}\cdot |\varphi''| \enspace \le \enspace (N+1)^{f-1}\cdot (N+1) \cdot |\varphi| \enspace
= \enspace (N+1)^{f}\cdot |\varphi|
\end{displaymath}
for the size of the formula $\varphi'$ without any fixpoint subformulas.
\qed

\subsection{Reachability Games}

\begin{defi}
A \emph{reachability game} between players $\exists$ and $\forall$ is a pointed and directed graph 
$\mathcal{G} = (V_\exists,V_\forall,E,v_0,W_\exists,W_\forall)$ with node set $V := V_\exists \cup V_\forall \cup W_\exists \cup W_\forall$
for some mutually disjoint $V_\exists,V_\forall,W_\exists,W_\forall$, edge relation $E \subseteq (V_\exists \cup V_\forall) \times V$ 
and designated starting node $v_0 \in V$.
Define $|\mathcal{G}| := |E|$ as the size of the game.

The sets $V_\exists$ and $V_\forall$ contain those nodes in which player $\exists$, resp.\ player $\forall$ makes a choice. The
sets $W_\exists$ and $W_\forall$ are terminal nodes in which player $\exists$, resp.\ player $\forall$ wins. We therefore require
that only nodes in $W_\exists$ or $W_\forall$ are terminal, i.e.\ for all $v \in V \setminus (W_\exists \cup W_\forall)$ there
is a $w \in V$ with $(v,w) \in E$.

A \emph{play} is a sequence $v_0,v_1,\ldots$ starting in $v_0$ and constructed as follows. If the play has visited nodes
$v_0,\ldots,v_i$ for some $i \in \Nat$ and $v_i \in V_p$ for some $p \in \{\exists,\forall\}$ then player $p$ chooses a node 
$w$ s.t.\ $(v_i,w) \in E$ and $v_{i+1} := w$. 

A play $v_0,\ldots,v_n$ is won by player $p$ if $v_n \in W_p$. A reachability game is called \emph{determined} if every play has a
unique winner. Given the prerequisite $W_\exists \cap W_\forall = \emptyset$, determinacy of a reachability game simply means that infinite plays are not possible.

A strategy\footnote{Here we restrict ourselves to memory-less strategies which are well-known to suffice for reachability games.} 
for player $p$ is a function $\sigma: V_p \to V$. A play $v_0,\ldots,v_n$ \emph{conforms} to a strategy $\sigma$ for player $p$
if for all $i=0,\ldots,n-1$ with $v_i \in V_p$: $v_{i+1} = \sigma(v_i)$. Such a strategy $\sigma$ is called \emph{winning strategy} if
player $p$ wins every play that conforms to $\sigma$. 

The problem of solving a determined reachability game is: given such a game $\mathcal{G}$, decide whether or not player $\exists$ has a 
winning strategy for $\mathcal{G}$.
\end{defi}

It is well-known that reachability games can be solved in linear time using dynamic programming for instance \cite{zermelo13}.

\begin{thm}
\label{thm:solvegame}
Solving a reachability game $\mathcal{G}$ can be done in time $O(|\mathcal{G}|)$.
\end{thm}

\subsection{Model Checking Games for Fixpoint-Free HFL}

In this section we define reachability games that capture exactly the satisfaction relation
for fixpoint-free \HFL formulas. 

Let $\varphi_0$ be a closed and fixpoint-free \HFL formula of type $\Pr$ and 
$\Transsys = (\States,\{\Transition{}{a}{} \mid a \in \Act\},\Labelling)$ a labeled transition system
with a designated starting state $s_0 \in \States$. The game $\game{\Transsys}{s_0}{\varphi_0}$
is played between players $\exists$ and $\forall$ in order to determine whether or not
$\Transsys, s_0 \models \varphi_0$ holds. A configuration of the game is written
\begin{displaymath}
s, f_1,\ldots,f_k, \eta \vdash \psi
\end{displaymath}
s.t.\ $\psi \in \fl{\varphi_0}$ is of some type $\tau_1 \to \ldots \to \tau_k \to \Pr$, $s \in \States$, 
and $f_i \in \StateTrans{\tau_i}$ for all $i = 1,\ldots,k$.
Note that $k = 0$ is possible. Finally, $\eta$ is a (partial) finite map that
assigns an element $f \in \StateTrans{\tau}$ to each
free variable $X$ of type $\tau$ in $\psi$ .

The intended meaning of such a configuration is: player $\exists$ tries to show 
$s \in \semTphiEta{\psi}\; f_1\ldots f_n$ whereas player $\forall$ tries to show the opposite.
Since the semantics of formulas is defined recursively, the play usually proceeds from one such
configuration to another containing a direct subformula. For instance, if the formula in the
current configuration is a disjunction then player $\exists$ chooses one of the disjuncts because
disjunctions are easy to prove but hard to refute in this way. Consequently, player $\forall$
performs a choice on conjunctions (negated disjunctions). A similar argument applies to 
configurations with modal operators. In case of function application we employ a small protocol 
of choices between these two players which simply reflects the semantics of function application 
in higher-order logic, etc.

A \emph{play} of $\game{\Transsys}{s_0}{\varphi_0}$ is a finite sequence $C_0,C_1,\ldots$
of configurations constructed as follows. $C_0 := s_0,\eta_0 \vdash \varphi_0$ where $\eta_0$ is
undefined on all arguments.

If $C_0,\ldots,C_{n-1}$ have already been constructed, then $C_n$ is obtained
by case distinction on $C_{n-1}$. 
\begin{enumerate}
\item \label{rule:disj} If $C_{n-1} = s, \eta \vdash \psi_1 \vee \psi_2$ then player $\exists$ chooses an $i \in \{1,2\}$
      and $C_{n} := s, \eta \vdash \psi_i$.

\item \label{rule:conj} If $C_{n-1} = s, \eta \vdash \neg(\psi_1 \vee \psi_2)$ then player $\forall$ chooses an $i \in \{1,2\}$
      and $C_{n} := s, \eta \vdash \neg\psi_i$.

\item \label{rule:diam} If $C_{n-1} = s, \eta \vdash \Mudiam{a}\psi$ then player $\exists$ chooses a $t \in \States$ s.t.\ 
      $\Transition{s}{a}{t}$ and $C_{n} := t, \eta \vdash \psi$.

\item \label{rule:box} If $C_{n-1} = s, \eta \vdash \neg\Mudiam{a}\psi$ then player $\forall$ chooses a $t \in \States$ s.t.\ 
      $\Transition{s}{a}{t}$ and $C_{n} := t, \eta \vdash \neg\psi$.

\item \label{rule:nneg} If $C_{n-1} = s, f_1,\ldots,f_k, \eta \vdash \neg\neg\psi$ then 
      $C_n := s, f_1,\ldots,f_k, \eta \vdash \psi$.

\item \label{rule:comp} If $C_{n-1} = s, f_1,\ldots,f_k, \eta \vdash \varphi\;\psi$ and $\psi$ is of type 
      $\sigma$ then player $\exists$ chooses a $g \in \StateTrans{\sigma}$. Next player $\forall$ has two options.
      \begin{itemize}
        \item He either continues with $C_n := s, g, f_1,\ldots,f_k, \eta \vdash \varphi$.
        \item Or let $\sigma = \sigma_1 \to \ldots \to \sigma_m \to \Pr$. Player $\forall$ chooses values 
              $h_i \in \StateTrans{\sigma_i}$ for $i=1,\ldots,m$, and either
              \begin{itemize}
                \item selects a $t \in g\; h_1\ldots h_m$, and the play continues with $t, h_1,\ldots,h_m, \eta \vdash \psi$, or
                \item selects a $t \not\in g\; h_1\ldots h_m$, and the play continues with $t, h_1,\ldots,h_m, \eta \vdash \neg\psi$.
              \end{itemize}
      \end{itemize}

\item \label{rule:ncomp} If $C_{n-1} = s, f_1,\ldots,f_k, \eta \vdash \neg(\varphi\;\psi)$ and $\psi$ is of type 
      $\sigma$ then player $\exists$ chooses a $g \in \StateTrans{\sigma}$. Next player $\forall$ has two options.
      \begin{itemize}
        \item He either continues with $C_n := s, g, f_1,\ldots,f_k, \eta \vdash \neg\varphi$.
        \item Or let $\sigma = \sigma_1 \to \ldots \to \sigma_m \to \Pr$. Player $\forall$ chooses values 
              $h_i \in \StateTrans{\sigma_i}$ for $i=1,\ldots,m$, and either
              \begin{itemize}
                \item selects a $t \in g\; h_1\ldots h_m$, and the play continues with $t, h_1,\ldots,h_m, \eta \vdash \psi$, or
                \item selects a $t \not\in g\; h_1\ldots h_m$, and the play continues with $t, h_1,\ldots,h_m, \eta \vdash \neg\psi$.
              \end{itemize}
      \end{itemize}

\item \label{rule:lambda} If $C_{n-1} = s, f_1,\ldots,f_k, \eta \vdash \lambda X.\psi$ then 
      $C_n := s, f_2,\ldots,f_k, \eta[X \mapsto f_1] \vdash \psi$.

\item \label{rule:nlambda} If $C_{n-1} = s, f_1,\ldots,f_k, \eta \vdash \neg\lambda X.\psi$ then 
      $C_n := s, f_2,\ldots,f_k, \eta[X \mapsto f_1] \vdash \neg\psi$.

\end{enumerate}
The game rules (\ref{rule:nneg}),(\ref{rule:lambda}) and (\ref{rule:nlambda}) are deterministic. Neither player has 
to make a real choice there.

A play $C_0,C_1,\ldots,C_n$ is won by player $\exists$, if
\begin{enumerate}
\item \label{wc:e1} $C_n = s, \eta \vdash q$ and $s \in \Labelling(q)$, or
\item \label{wc:e2} $C_n = s, \eta \vdash \neg q$ and $s \not\in \Labelling(q)$, or
\item \label{wc:e3} $C_n = s, f_1,\ldots,f_k,\eta \vdash X$ and $s \in \eta(X)\;f_1\ldots f_k$, or
\item \label{wc:e4} $C_n = s, f_1,\ldots,f_k,\eta \vdash \neg X$ and $s \not\in \eta(X)\;f_1\ldots f_k$, or
\item \label{wc:e5} $C_n = s, \eta \vdash \neg\Mudiam{a}\psi$ and there is no $t \in \States$ with $\Transition{s}{a}{t}$.
\end{enumerate} 
Player $\forall$ wins this play, if
\begin{enumerate}
\setcounter{enumi}{5}
\item \label{wc:a1} $C_n = s, \eta \vdash q$ and $s \not\in \Labelling(q)$, or
\item \label{wc:a2} $C_n = s, \eta \vdash \neg q$ and $s \in \Labelling(q)$, or
\item \label{wc:a3} $C_n = s, f_1,\ldots,f_k,\eta \vdash X$ and $s \not\in \eta(X)\;f_1\ldots f_k$, or
\item \label{wc:a4} $C_n = s, f_1,\ldots,f_k,\eta \vdash \neg X$ and $s \in \eta(X)\;f_1\ldots f_k$, or
\item \label{wc:a5} $C_n = s, \eta \vdash \Mudiam{a}\psi$ and there is no $t \in \States$ with $\Transition{s}{a}{t}$.
\end{enumerate} 

We remark that these games do not easily extend to formulas with fixpoint quantifiers and variables via the
characterisation of the model checking problem for the modal $\mu$-calculus as a parity game \cite{Stirling95}.
The natural extension would add simple unfolding rules for fixpoint constructs which lead to infinite plays.
The type of the outermost fixpoint variable that gets unfolded infinitely often in such a play would determine the
winner. 

However, this is neither sound nor complete. Consider the formula $(\nu (X: \Pr \to \Pr).\mu (Y:\Pr).X\;Y)\;\false$.
It is equivalent to $\true$, hence, player $\exists$ should have a winning strategy for the game on this formula
and any transition system. But player $\forall$ can enforce a play via rule (\ref{rule:comp}) in which the outermost
variable that gets unfolded infinitely often is $Y$ which is of type $\mu$. 

This shows that the straight-forward extension to non-fixpoint-free formulas is not complete. Because of the presence
of negation it is also not sound. Another explanation for the failure of such games is given by the model checking
games for FLC \cite{lange:ah-flc:06} which incorporate a stack and a visibly pushdown winning condition in order to 
model that the variable $X$ (the function) is more important than the variable $Y$ (the argument) in the example 
above.

\begin{lem}
\label{lem:winplay}
Every play of $\game{\Transsys}{s}{\varphi}$ has a unique winner.
\end{lem}

\proof
All rules properly reduce the size of the formula component in a configuration. Hence, there are no infinite plays, and a play
is finished when either one of the players cannot perform a choice or there is no rule that applies to the current
configuration anymore. 

Note that for as long as rules still apply there are only two situations in
which a player can get stuck: Either the current
configuration is $s, \eta \vdash \Mudiam{a}\psi$ or it is $s, \eta \vdash \neg\Mudiam{a}\psi$ and there is no $t \in \States$ s.t.\ 
$\Transition{s}{a}{t}$. These cases are covered by winning conditions (\ref{wc:e5}) and (\ref{wc:a5}).

All other rules always guarantee one player a possible choice. The only rules for which this is not obvious are (\ref{rule:comp})
and (\ref{rule:ncomp}). First note that $\semb{\sigma}^\Transsys$ is non-empty for any type $\sigma$. Hence, player $\exists$
can always choose some $g$. Then let, for some arguments $h_1,\ldots,h_m$ chosen by player $\forall$, $T := g\; h_1\ldots h_m$.
Note that it is impossible to have $T = \emptyset$ and at the same time $\States\setminus T = \emptyset$ for as long as 
$\States \ne \emptyset$ for the underlying
state space $\States$. Hence, player $\forall$ cannot get stuck in this rule either.

If a play finishes because no rule applies then the formula in the current configuration must either be atomic or a negation
of an atomic formula, i.e.\ of one of the forms $q,\neg q, X, \neg X$ for some $q \in \Prop$, $X \in \Var$. In any case, one
of the winning conditions (\ref{wc:e1})--(\ref{wc:e4}) and (\ref{wc:a1})--(\ref{wc:a4}) applies.

This shows that every play has at least one winner. Finally, it is not hard to see that the winning conditions are mutually
exclusive, i.e.\ every play has at most one winner.
\qed

\begin{thm}
\label{thm:gamescomplete}
Let $\varphi_0$ be closed, fixpoint-free, and of type $\Pr$. If $s_0 \in \semb{\varphi_0}^\Transsys$ then player 
$\exists$ has a winning strategy for the game $\game{\Transsys}{s_0}{\varphi_0}$.
\end{thm}

\proof
We call a configuration $C = t, f_1,\ldots,f_k, \eta \vdash \psi$ of the game $\game{\Transsys}{s_0}{\varphi_0}$ \emph{true} 
if $t \in \semb{\psi}^\Transsys_{\eta}\; f_1\ldots f_k$. Otherwise we call $C$ \emph{false}.

Suppose $s_0 \in \semb{\varphi_0}$, i.e. the starting configuration $s_0, \eta_0 \vdash \varphi_0$ of $\game{\Transsys}{s_0}{\varphi_0}$
is true. Player $\exists$'s strategy will consist of preservering truth along a play. We will
show by case distinction on the last rule played that player $\exists$ can enforce a play in which every configuration is true.
I.e.\ if a configuration that is true requires her to make a choice then she can
choose a successor configuration which is also true. If such a configuration requires player $\forall$ to make a choice
then regardless of what he selects, the successor will always be true.

\paragraph{Cases (\ref{rule:disj}) and (\ref{rule:conj}), the Boolean operators.} If a play has reached a configuration 
$t, \eta \vdash \psi_1 \vee \psi_2$ that is true then there is an $i \in \{1,2\}$ s.t.\ $t, \eta \vdash \psi_i$ is true. 
Player $\exists$  chooses this $i$. Note that player $\forall$ will ultimately preserve truth if he makes a choice in a configuration 
$t, \eta \vdash \neg(\psi_1 \vee \psi_2)$.

\paragraph{Cases (\ref{rule:diam}) and (\ref{rule:box}), the modal operators.} Similarly, player $\exists$ can preserve truth 
in a configuration of the form $t, \eta \vdash \Mudiam{a}\psi$, and player $\forall$ must preserve truth in a configuration of the form 
$t, \eta \vdash \neg\Mudiam{a}\psi$.

\paragraph{Case (\ref{rule:nneg}), double negation.} Preservation of truth is trivial.

\paragraph{Case (\ref{rule:comp}), positive application.} Suppose the play has reached a configuration 
$t, f_1,\ldots,f_k,\eta \vdash \varphi\; \psi$ that is
true. Let $g := \semb{\psi}^\Transsys_{\eta}$. Note that $g$ always exists, hence, 
player $\exists$ can choose it. By $\beta$-equivalence we have
\begin{displaymath}
t \in \semb{\varphi\;\psi}^\Transsys_{\eta}\; f_1\ldots f_k \enspace \Rightarrow \enspace 
t \in \semb{\varphi}^\Transsys_{\eta}\;g\;f_1\ldots f_k
\end{displaymath}
which shows that truth is preserved if player $\forall$ selects his first option.

Suppose he selects his second option with arguments $h_1,\ldots,h_m$ for $g$ instead. Since $g = \semb{\psi}^\Transsys_{\eta}$
we obviously have for all $t \in \States$: $t \in g\; h_1\ldots h_m$ iff $t \in \semb{\psi}^{\Transsys}_{\eta}\; h_1\ldots h_m$. 
This shows that truth is preserved regardless of which way player $\forall$ leads.

\paragraph{Case (\ref{rule:ncomp}), negative application.} This is the same as the case above. Note that -- by the semantics of the 
negation operator -- we have $\neg(\varphi\; \psi) \equiv (\neg\varphi)\; \psi$. 

\paragraph{Cases (\ref{rule:lambda}) and (\ref{rule:nlambda}), $\lambda$-abstraction.} This is only an equivalence-preserving $\beta$-reduction. 
Hence, truth is preserved. For case (\ref{rule:nlambda}) remember that the complement of a function is defined pointwise.
\vskip3mm

It remains to be seen that this truth-preserving strategy guarantees player $\exists$ to win any play. I.e.\ 
assume that player $\forall$ uses his best strategy against player $\exists$'s truth-preserving strategy and consider the unique
play $C_0,\ldots,C_n$ that results from playing against each other. By the argumentation above, we know that $C_i$ is true for all
$i=0,\ldots,n$. A quick inspection of player $\forall$'s winning conditions (\ref{wc:a1})--(\ref{wc:a5}) shows that he cannot be the
winner of this play because all of them require the play at hand to end in a configuration that is not true. 

According to Lemma~\ref{lem:winplay}, player $\exists$ wins every play in which she uses the truth-preserving strategy. Hence, this
is a winning strategy. 
\qed

\begin{thm}
\label{thm:gamessound}
Let $\varphi_0$ be closed, fixpoint-free, and of type $\Pr$. If $s_0 \not\in \semb{\varphi_0}^\Transsys$ then 
player $\forall$ has a winning strategy for the game $\game{\Transsys}{s_0}{\varphi_0}$.
\end{thm}

\proof
Similar to the proof of Theorem~\ref{thm:gamescomplete}. The starting configuration of $\game{\Transsys}{s_0}{\varphi_0}$ is false. 
An analysis of the game rules shows that player $\forall$ can preserve falsity with his choices, and player $\exists$ must preserve
falsity. 

This is shown for rules (\ref{rule:disj}), (\ref{rule:conj}), (\ref{rule:diam}), (\ref{rule:box}) and (\ref{rule:nneg}) in the same
way as above in the proof of Thm.~\ref{thm:gamescomplete}. Here we only consider the case (\ref{rule:comp}). The case of rule
(\ref{rule:ncomp}) is shown analogously.

Suppose the current configuration is $s, f_1,\ldots,f_k, \eta \vdash \varphi\; \psi$, s.t.\ $\psi$ has type $\sigma$ and player
$\exists$ has chosen some $g \in \semb{\sigma}^\Transsys$. We need to distinguish two subcases.

If $s \not\in \semTphiEta{\varphi}\; g\; f_1\ldots f_k$ then player $\forall$ can easily preserve falsity by choosing the successor
configuration $s, g,f_1,\ldots,f_k,\eta \vdash \varphi$.

If $s \in \semTphiEta{\varphi}\; g\; f_1\ldots f_k$ then we must have $g \ne_{\sigma} \semTphiEta{\psi}$ for equality would,
by $\beta$-equivalence, contradict the assumption that the current configuration is false. Remember that $\ne_{\sigma}$ is
inequality w.r.t.\ the pointwise order $\sqsubseteq_\sigma$. Now suppose $\sigma = \sigma_1 \to \ldots \to \sigma_m \to \Pr$.
Hence, there must be $h_i \in \semb{\sigma_i}^\Transsys$ for $i = 1,\ldots,m$ s.t.\ $g\; h_1\ldots h_m \ne \semTphiEta{\psi}\; h_1\ldots h_m$.
First of all, player $\forall$ can choose these arguments $h_1,\ldots,h_m$. Next, let $T := g\; h_1\ldots h_m$ and 
$T' := \semTphiEta{\psi}\; h_1\ldots h_m$. Note that $T,T' \subseteq \States$. Hence, $T \ne T'$ means $T \not\subseteq T'$ 
or $T' \not\subseteq T$.

In the first case there is a $t \in g\; h_1\ldots h_m$ s.t.\ $t \not\in \semTphiEta{\psi}\; h_1\ldots h_m$. Player $\forall$ can choose 
this $t$ and continue with the configuration $t, h_1,\ldots,h_m, \eta \vdash \psi$ which is false.

In the second case note that $T' \not\subseteq T$ iff $\States \setminus T \not\subseteq \States \setminus T'$.
Hence, there is a $t \not\in g\; h_1\ldots h_m$ s.t.\ $t \not\in \States \setminus (\semTphiEta{\psi}\; h_1\ldots h_m)$. Again,
player $\forall$ can choose this $t$ and continue with the false configuration $t, h_1,\ldots,h_m, \eta \vdash \neg\psi$.

The proof is finished just like the proof of Thm.~\ref{thm:gamescomplete}. With this strategy, player $\forall$ can always enforce
a play that ends in a false configuration, but player $\exists$ can only win plays that end in true configurations.
\qed

Putting these two theorems together shows that these games correctly characterise the satisfaction relation for fixpoint-free \HFL.

\begin{cor}
\label{cor:gamescorrect}
For all transition systems $\Transsys$ all of their states $s$, and all fixpoint-free \HFL formulas $\varphi$ of type $\Pr$ we have:
$\Transsys, s \models \varphi$ iff player $\exists$ wins the game $\game{\Transsys}{s}{\varphi}$.
\end{cor}

\begin{lem}
\label{lem:mcgame}
For any $k,m \ge 1$, any $\Transsys = (\States,\{\Transition{}{a}{} \mid a \in \Act\},\Labelling)$ with 
$|\States| = n$, any $s \in \States$, and
any \hfl{k,m} formula $\varphi$, $\game{\Transsys}{s}{\varphi}$ is a reachability game of size at most
\begin{displaymath}
4n^2\cdot|\varphi|^2\cdot(m^{(k-1)m^{k-2}}\cdot\tower{n(k-1+m)^{k-1}}{k})^{2(m+v(\varphi))}\ .
\end{displaymath}
\end{lem}

\proof
It should be clear from the definition of the game that $\game{\Transsys}{s}{\varphi}$ can indeed be regarded as a 
reachability game $(V_\exists,V_\forall,v_0,W_\exists,W_\forall)$. Its node set $V := V_\exists \cup V_\forall \cup W_\exists \cup W_\forall$ 
consists of all possible 
configurations in $\game{\Transsys}{s}{\varphi}$ plus auxiliary configurations that represent the choices done by either
player in rules (\ref{rule:comp}) and (\ref{rule:ncomp}) which require an alternating sequence of choices of fixed depth 3.
However, this can at most double the number of nodes in comparison to the number of configurations. 

The starting node $v_0$ is the starting configuration $s, \eta \vdash \varphi$
for an everywhere undefined $\eta$.
The partition of the nodes is given by the definition of the game rules and winning conditions above: $V_\exists$, resp.\ 
$V_\forall$ are all those configurations that require player $\exists$, resp.\ $\forall$ to make a choice -- including
the auxiliary configurations for the choices in between rules. The edges of the game are simply given by the game rules.
$W_\exists$, resp.\ $W_\forall$ are all those configurations that end a play according to one of the winning conditions. 
Lemma~\ref{lem:winplay} shows that these games are determined.

What remains to be seen is that the size of $\game{\Transsys}{s}{\varphi}$ is bounded accordingly.
There are at most $n$ different states $t \in \States$, and at most $|\varphi|$ many formulas $\psi \in \fl{\varphi}$. 
The maximal width of a configuration, the parameter $m'$ in $t, f_1, \ldots,
f_{m'}, \eta \vdash \psi$ is bounded by $m$ since here $\psi$ has a type of arity $m'$. According to Lemma~\ref{lem:sizeofalltypes} there 
are at most $m^{(k-1)m^{k-2}}\tower{n(k-1+m)^{k-1}}{k}$ many different functions $f_i$ of type order $k-1$. None of these can be of type
order $k$ because they only occur as arguments to formulas of strictly higher order. We simply define $m^{(k-1)m^{k-2}} := m$ if
$k = 1$ rather than introducing $\max$-operators in these terms.

Finally, we need to estimate the number of different environments $\eta$. These map at most each $\lambda$-bound
variable $X$ of type $\tau$ in $\varphi$ to an element of $\StateTrans{\tau}$. Again, if $X$ occurs bound in
$\varphi$, then there is a $\lambda X.\psi \in \fl{\varphi}$ of type $\sigma$, and we have $\order{\sigma} \ge \order{\tau}+1$.
Hence, there are at most $m^{(k-1)m^{k-2}}\tower{n(k-1+m)^{k-1}}{k}$ many possible values for each such $X$, and thus at most
$(m^{(k-1)m^{k-2}}\tower{n(k-1+m)^{k-1}}{k})^{v(\varphi)}$ many different environments $\eta$.

Putting this together we obtain
\begin{align*}
&2 \cdot n \cdot (m^{(k-1)m^{k-2}}\cdot\tower{n(k-1+m)^{k-1}}{k})^m \cdot (m^{(k-1)m^{k-2}}\cdot\tower{n(k-1+m)^{k-1}}{k})^{v(\varphi)}\cdot |\varphi| 
\enspace = \\
& 2n\cdot|\varphi|\cdot(m^{(k-1)m^{k-2}}\cdot\tower{n(k-1+m)^{k-1}}{k})^{m+v(\varphi)}
\end{align*}
as an upper bound on the number of nodes in $\game{\Transsys}{s}{\varphi}$. The number of edges in this directed graph
can of course be at most quadratic in the number of nodes which finishes the proof. \qed

\subsection{The Model Checking Complexity}

\begin{thm}
The model checking problem on a transition system $\Transsys$ of size $n$ and an \hfl{k,m} formula $\varphi$ can be 
solved in time $2^{O(|\varphi|^k\cdot \log (n\cdot |\varphi|))} \cdot (\tower{n(m+k-1)^{k-1}}{k})^{O(|\varphi|^2)}$ for any $k,m \ge 1$.
\end{thm}

\proof
Let $\States$ be the state space of $\Transsys$, $n := |\States|$, and $\varphi \in$ \hfl{k,m} for some $k,m \ge 1$ be closed.
According to Lemma~\ref{lem:fpelimination} there is a fixpoint-free $\varphi'$ s.t.\ 
\begin{itemize}
\item $v(\varphi') \le v(\varphi) + m$, 
\item $|\varphi'| \le |\varphi|\cdot(n+1)^{|\varphi|}\cdot (\tower{n(m+k-1)^{k-1}}{k})^{m\cdot |\varphi|}$, and
\item for all $s \in \States$ we have $\Transsys, s \models \varphi$ iff $\Transsys, s \models \varphi'$.
\end{itemize}
Now take any $s \in \States$. Consider the reachability game $\game{\Transsys}{s}{\varphi'}$. According to Lemma~\ref{lem:mcgame} 
its size is at most
\begin{align*}
& 4n^2\cdot|\varphi'|^2\cdot(m^{(k-1)m^{k-2}}\cdot\tower{n(k-1+m)^{k-1}}{k})^{2(m+v(\varphi'))} \\
= \enspace & 4n^2\cdot (|\varphi|\cdot(n+1)^{|\varphi|}\cdot
(\tower{n(m+k-1)^{k-1}}{k})^{m\cdot
  |\varphi|}){}^2\cdot(m^{(k-1)m^{k-2}}\cdot\tower{n(k-1+m)^{k-1}}{k})^{2(2m+v(\varphi))}
\end{align*}
by replacing $|\varphi'|$ according to Lemma \ref{lem:fpelimination}. This can
be approximated from above by
\begin{align*}
& 4 \cdot (n+1)^{2|\varphi|+2} \cdot |\varphi|^{3|\varphi|^{k}+2} \cdot (\tower{n(m+k-1)^{k-1}}{k})^{2|\varphi|^2 + 6|\varphi|} \\
= \enspace & n^{O(|\varphi|)} \cdot |\varphi|^{O(|\varphi|^{k})} \cdot (\tower{n(m+k-1)^{k-1}}{k})^{O(|\varphi|^2)}) 
\enspace = \enspace 2^{O(|\varphi|^k\cdot \log (n\cdot |\varphi|))} \cdot (\tower{n(m+k-1)^{k-1}}{k})^{O(|\varphi|^2)}
\end{align*}
because $|\varphi|$ is an upper bound on $m$, $k$ and $v(\varphi)$. By Cor.~\ref{cor:gamescorrect} we have
$\Transsys, s \models \varphi'$ iff player $\exists$ has a winning strategy for $\game{\Transsys}{s}{\varphi'}$.
And by Thm.~\ref{thm:solvegame} the asymptotic time needed to solve this game equals its size.
\qed

\begin{cor}
\label{cor:inkexptime}
For any $k,m \ge 1$ the \hfl{k,m} model checking problem is in \EXPTIME{k}.
\end{cor}

\begin{cor}
\label{cor:exprcomplupper}
For any $k,m \ge 1$ the model checking problem for \hfl{k,m} on a fixed transition system is in \ExpTime.
\end{cor}

\proof
If $k,m$ and $n$ are fixed constants then so is $\tower{n(k-1+m)^{k-1}}{k}$. Hence, model checking in this case can be done
in time $2^{O(|\varphi|^2 + |\varphi|^k\cdot \log |\varphi|)}$. 
\qed


%% file: lower.tex
\section{The Lower Bound}
\label{sec:lower}

We will show that the upper bound in Cor.~\ref{cor:inkexptime} is optimal by reducing the word problem for alternating space bounded
Turing Machines to the model checking problem for \hfl{}. 

Let $F_0(p(n)) := 2^{p(n)}$ and $F_{k+1}(p(n)) := 2^{p(n) \cdot F_{k}(p(n))}$ for any polynomial $p(n)$. A simple 
induction shows 
$F_k(p(n)) \ge \tower{p(n)}{k+1}$ for all $k,n \in \Nat$. Clearly, the space used by a $\tower{p(n)}{k}$-space bounded
Turing Machine is also bounded by $F_{k-1}(p(n))$ for $k \ge 1$. This slight shift in indices makes the encoding of
large numbers in the next section easier. On the other hand, it only allows us to consider alternating $\tower{p(n)}{k}$-space
bounded Turing Machines when $k \ge 1$. Hence, we will only obtain \EXPTIME{k}-hardness results for $k \ge 2$.
Fortunately, the results for the \hfl{1,m} fragments follow from known lower bounds for \FLC \cite{lange:3mcflc:05}.

\subsection{Representing Large Numbers in \HFL}
\label{repres}

Let $\tau_0 := \Pr$ and $\tau_{k+1} := \tau_k \to \Pr$ for all $k \in \Nat$. Note that on a transition system of exactly $p(n)$ states 
we have $|\tau_k| = F_{k}(p(n))$ for all $k \in \Nat$. 
In order to model the position of the head and the sequence of the cells of a tape of size $F_k(p(n))$ we therefore use a transition
system $\Transsys$ with $p(n)$ many states, and an encoding of the natural numbers $\{0,\ldots,F_k(p(n))-1\}$ via \HFL 
functions\footnote{We will also use the term ``function'' for an object of type $\tau_0$ which is a set strictly speaking, hence,
a function of order 0.} of type $\tau_k$ over $\Transsys$. This is done by induction on $k$. Let $n$ and the polynomial $p(n)$ be fixed.

For $k = 0$ we assume that $\Transsys$ contains $p(n)$ many states called $0,\ldots,p(n)-1$.
A number $i$ between $0$ and $F_0(p(n))-1$ is now represented by the subset $S_i = \{j \mid$ the $j$-th bit of $i$ is $1\}$ 
which has type $\tau_0$.
Let $\repr{i}{0}$ for $i \in \{0,\ldots,F_0(p(n))-1\}$ denote the function of type $\tau_0$ that represents the natural number $i$ in this way.

Now let $k > 0$. By assumption there are \HFL functions $\repr{0}{k-1},\ldots,$ $\repr{F_{k-1}(p(n))-1}{k-1}$ of type 
$\tau_{k-1}$  that represent the numbers $0,\ldots,F_{k-1}(p(n))-1$. Clearly, these are linearly ordered by the standard ordering on the numbers
that they represent. We now need to find a representation of the numbers $0,\ldots,F_k(p(n))-1$ via \HFL functions of type
$\tau_k = \tau_{k-1} \to \Pr$. 

\begin{figure}[t]
\begin{center}
\begin{math}
\begin{array}{c|ccccc}
\rule[-3mm]{0pt}{8mm}  & \enspace\repr{F_{k-1}(p(n))-1}{k-1}\enspace & \ldots & \enspace \repr{1}{k-1}\enspace & \enspace \repr{0}{k-1}\enspace                \\ \hline\hline
\rule[-3mm]{0pt}{8mm} \repr{0}{k} & \repr{0}{0} & \ldots & \repr{0}{0} &  \repr{0}{0}                      \\ \hline
\rule[-3mm]{0pt}{8mm} \repr{1}{k} & \repr{0}{0} & \ldots & \repr{0}{0} &  \repr{1}{0}                      \\ \hline
\rule[-3mm]{0pt}{8mm} \repr{2}{k} & \repr{0}{0} & \ldots & \repr{0}{0} &  \repr{2}{0}                      \\ \hline
\rule[-3mm]{0pt}{8mm}  \vdots & \vdots          & \vdots &  \vdots & \vdots                                           \\ \hline
\rule[-3mm]{0pt}{8mm} \repr{F_1(p(n))-1}{k} & \repr{0}{0} &  \ldots & \repr{0}{0} &  \repr{F_0(p(n))-1}{0}            \\ \hline
\rule[-3mm]{0pt}{8mm} \repr{F_1(p(n))}{k} & \repr{0}{0} &  \ldots & \repr{1}{0} & \repr{0}{0}                     \\ \hline
\rule[-3mm]{0pt}{8mm} \repr{F_1(p(n))+1}{k} & \repr{0}{0} &  \ldots & \repr{1}{0} & \repr{1}{0}                     \\ \hline
\rule[-3mm]{0pt}{8mm} \vdots & \vdots            & \vdots &  \vdots & \vdots                                           \\ \hline
\rule[-3mm]{0pt}{8mm} \repr{F_k(p(n))-1}{k} & \repr{F_0(p(n))-1}{0} & \ldots & \enspace \repr{F_0(p(n))-1}{0}\enspace & \enspace \repr{F_0(p(n))-1}{0} \enspace             \\
\hline
\end{array}
\end{math}
\end{center}

\caption{Encoding large numbers as lexicographically ordered functions.}
\label{fig:encode}
\end{figure}

These functions have a finite and linearly ordered domain as well as co-domain. Hence, we can regard them as lexicographically
ordered words of length $F_{k-1}(p(n))$ over the alphabet
$\{\repr{0}{0},\ldots,\repr{F_0(p(n))-1}{0}\}$, or simply as base-$F_0(p(n))$
numerals with $F_{k-1}(p(n))$ digits. Now $\repr{i}{k}$ simply is the 
$i$-th function in this lexicographic ordering as depicted in Fig.~\ref{fig:encode}. 
The leftmost column contains the symbolic name $\repr{i}{k}$ for the $i$-th function in that ordering. The upper row contains the ordered
list of all possible arguments $x$ for any such $\repr{i}{k}$ while the entries below denote the values $\repr{i}{k}\; x$.

\subsection{The Reduction}

For the remainder of this section we fix an alternating $F_k(p(n))$-space bounded Turing Machine 
$\mathcal{M} = (Q,\Sigma,\Gamma,q_0,\delta,q_{\mathit{acc}},q_{\mathit{rej}})$ and an input word $w$ of length $n$. W.l.o.g. we 
assume $p(n) > n$ for all $n \in \Nat$.
According to Thm.~\ref{thm:alternation} we can also assume $\Sigma = \Gamma$ and $|\Gamma| = 2$. 

Of course, symbols are just purely 
syntactic objects. However, later we need to encode these two symbols as propositions in transition systems, and we will use the 
propositions $\true$ and $\false$ to do so. Hence, we can simplify notation slightly by assuming $\Gamma = \{\true,\false\}$ 
as two different alphabet letters with no attached meaning. W.l.o.g.\ we assume that the special blank symbol $\Box$ is encoded
by a sequence of the symbol $\false$ of some suitable length.

The goal is to construct a transition system $\Transsys_{\mathcal{M},w}$ and an \hfl{k+2} formula 
$\Phi^k_{\mathcal{M},w}$ both of polynomial size, s.t.\ $\Transsys_{\mathcal{M},w}, s \models \Phi^k_{\mathcal{M},w}$ iff 
$w \in L(\mathcal{M})$ for some state $s$. The types of the subformulas of $\Phi^k_{\mathcal{M},w}$ that we present in the following
can easily be inferred. We will therefore omit type annotations.

\begin{figure}[t]
\framebox[\textwidth]{
\parbox{\textwidth}{
\begin{center}
\scalebox{0.6}{\includegraphics{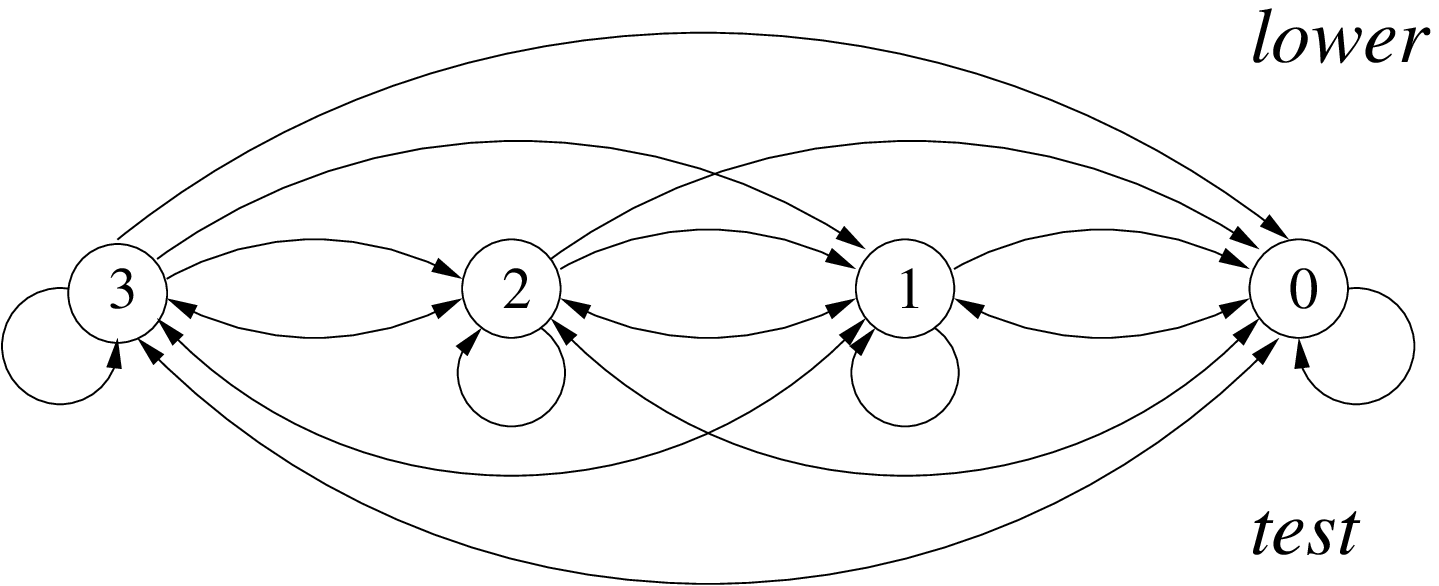}}
\end{center}
}}
\caption{The transition system $\Transsys_{\mathcal{M},w}$ for $p(n) = 4$.}
\label{fig:transsys1}
\end{figure}

We begin with the construction of the transition system.
Let $\Prop := \emptyset$, i.e.\ no state of $\Transsys_{\mathcal{M},w}$ carries a label. There are two modal accessibility relations
with labels $\mathit{lower}$ and $\mathit{test}$.
Let $\Transsys_{\mathcal{M},w} = (\States,\{\Transition{}{\mathit{lower}}{},\Transition{}{\mathit{test}}{}\},L)$ where $\States = \{0,\ldots,p(n)-1\}$, 
and $L$ maps every state to the empty set. The $\mathit{lower}$-relation simply resembles the less-than-relation on natural numbers: 
$\Transition{i}{\mathit{lower}}{j}$ iff $j < i$. The $\mathit{test}$-relation forms a clique: $\Transition{i}{\mathit{test}}{j}$ for all
$i,j \in \States$. It is used to form global statements. Note that for all states $i,j$, and all formulas $\psi$:
$i \models \Mubox{\mathit{test}}\psi$ iff $j \models \Mubox{\mathit{test}}\psi$. Fig.~\ref{fig:transsys1} depicts $\Transsys_{\mathcal{M},w}$
for $p(n) = 4$. The transitions above the states are the $\Transition{}{\mathit{lower}}{}$-relation. Consequently, they only lead from
the left to the right. The transitions below the states are the $\Transition{}{\mathit{test}}{}$-relation.

For the remainder of this section we fix $\Transsys_{\mathcal{M},w}$ as the transition system over which formulas are interpreted
and write $\semb{\cdot}$ instead of $\semb{\cdot}^{\Transsys_{\mathcal{M},w}}$.

Remember that any function of type $\tau_0$ represents a number in binary coding over $\Transsys_{\mathcal{M},w}$: 
$\repr{i}{0} = \{ j \mid$ the $j$-th bit of $i$ is $1 \}$. Furthermore, the transitions in $\Transsys_{\mathcal{M},w}$ allow
a bit to assess the values of all lower bits.
The \hfl{1,1} formula 
\begin{displaymath}
\mathit{inc}^0 \enspace := \enspace \lambda X. X \leftrightarrow \Mudiam{\mathit{lower}}\neg X
\end{displaymath}
models the increment among the number representations $\repr{0}{0},\ldots,\repr{F_0(p(n))-1}{0}$. Increment of a binary counter 
sets a bit of the input to $0$ if itself and all lower bits are $1$. A bit is set to $1$ if it currently is $0$ and all lower bits are $1$.
A bit is preserved if a lower-valued bit is unset. Applied to $\repr{F_0(p(n))-1}{0}$ this yields $\repr{0}{0}$ again. Similarly, we can
model the decrement among these values as
\begin{displaymath}
\mathit{dec}^0 \enspace := \enspace \lambda X. X \leftrightarrow \Mudiam{\mathit{lower}} X
\end{displaymath}

\begin{lem}
\label{lem:incdeccorrect0}
For all $i \in \{0,\ldots,F_0(p(n))-1\}$ we have:
\begin{itemize}
\item[a)] $\semb{\mathit{inc}^0}\; \repr{i}{0} \enspace = \enspace \repr{i+1 \mod F_0(p(n))}{0}$,
\item[b)] $\semb{\mathit{dec}^0}\; \repr{i}{0} \enspace = \enspace \repr{i-1 \mod F_0(p(n))}{0}$.
\end{itemize}
\end{lem}

\proof
We will only show part (a) since part (b) is entirely analogous. Take any $i \in \{0,\ldots,F_0(p(n))\}$ and let 
$m := p(n) - 1 = (\log F_0(p(n))) - 1$. Furthermore, let $b_m \ldots b_0 \in \{0,1\}^{m+1}$ be the binary representation of the number
$i$. According to the encoding described in the previous section, we have $\repr{i}{0} = \{ j \mid b_j = 1 \}$.

Now take any state $j$ and suppose $j \in \semb{\mathit{inc}^0}\; \repr{i}{0}$. The body of the $\lambda$-abstracted formula
$\mathit{inc}^0$ is a bi-implication which can be seen as an abbreviation of a disjunction of two conjunctions. Hence, there are two 
possibilities. 
\begin{itemize}
\item Either $j \models X \wedge \Mudiam{\mathit{lower}}\neg X$ with $X$ interpreted as $\repr{i}{0}$. This means that bit $j$ is set in $i$ and there
      is a lower bit that is not set in $i$. Hence, bit $j$ is also set in $i+1 \mod F_0(p(n))$.
\item Or $j \models \neg X \wedge \Mubox{\mathit{lower}} X$ under the same interpretation of $X$. Then the $j$-th bit is the lowest bit which is unset
      in $i$. Hence, it gets set in $i+1 \mod F_0(p(n))$.
\end{itemize}
This shows that only those bits are included in the increment process that should be included. The converse direction -- all necessary
bits are included -- is shown in the same way by case analysis. Suppose $j \not\in \semb{\mathit{inc}^0}\; \repr{i}{0}$, i.e.\ 
$j \not\models X \leftrightarrow \Mudiam{\mathit{lower}}\neg X$ with $X$ interpreted by $\repr{i}{0}$.
\begin{itemize}
\item Either $j \models X \wedge \Mubox{\mathit{lower}} X$. Then the $j$-th bit is among all those least bits that are set in $i$. Hence, it gets unset
      in $i+1 \mod F_0(p(n)).$
\item Or $j \models \neg X \wedge \Mudiam{\mathit{lower}} \neg X$. Then the $j$-th bit is not set in $i$, but there is a lower bit that is not set
      either. Hence, it preserves its value and remains unset in $i+1 \mod F_0(p(n))$.
\end{itemize}
\qed

In order to define the increment and decrement of numbers $\repr{i}{k}$ in lexicographic ordering for some $k > 0$ we need to have equality,
less-than and greater-than tests on lower types $\tau_{k-1}$. They can be implemented as functions of type $\tau_{k-1} \to \tau_{k-1} \to \Pr$.
Equality simply makes use of the fact that two numbers are equal iff they have the same binary representation.
\begin{displaymath}
\mathit{eq}^0 \enspace := \enspace \lambda I. \lambda J.\Mubox{\mathit{test}}(I \leftrightarrow J)
\end{displaymath}
The other comparing functions need to access single bits. Remember that $i < j$ iff there is a bit that is unset in $i$ and set in $j$
s.t.\ $i$ and $j$ agree on all higher bits. We therefore first define formulas $\mathit{bit}_i$ for $i=0,\ldots,p(n)-1$ s.t.\ $\mathit{bit}_i$ axiomatises
the state $i$, i.e.\ $j \models \mathit{bit}_i$ iff $i=j$. This can be done recursively as
\begin{displaymath}
\mathit{bit}_0 \enspace := \enspace \Mubox{\mathit{lower}}\false \enspace, \quad \enspace 
\mathit{bit}_{i+1} \enspace := \enspace \Mudiam{\mathit{lower}}\mathit{bit}_i \wedge \Mubox{\mathit{lower}}(\bigvee\limits_{j \le i} \mathit{bit}_j)
\end{displaymath}
Note that $|\mathit{bit}_i| = O(i^2)$ only. 
\begin{align*}
\mathit{lt}^0 \enspace &:= \enspace \lambda I. \lambda J.\enspace \bigvee\limits_{k=0}^{p(n)-1} \Mubox{\mathit{test}}
    \Big((\mathit{bit}_k \to \neg I \wedge J)) \wedge 
    \bigwedge\limits_{h>k} \big((\mathit{bit}_h \to (I \leftrightarrow J)\big)\Big) \\
\mathit{gt}^0 \enspace &:= \enspace \lambda I. \lambda J.(\mathit{lt}^0\; J\; I)
\end{align*}

\begin{lem} 
\label{lem:eqltgtcorrect0}
For all $i,j \in \{0,\ldots,F_0(p(n))-1\}$ we have:
\begin{itemize}
\item[a)] $\semb{\mathit{eq}^0}\; \repr{i}{0}\; \repr{j}{0} \enspace = \enspace 
\begin{cases}
\States , & \mbox{if } i = j \\
\emptyset , & \mbox{o.w.}
\end{cases}$

\item[b)] $\semb{\mathit{lt}^0}\; \repr{i}{0}\; \repr{j}{0} \enspace = \enspace 
\begin{cases}
\States , & \mbox{if } i < j \\
\emptyset , & \mbox{o.w.}
\end{cases}$

\item[c)] $\semb{\mathit{gt}^0}\; \repr{i}{0}\; \repr{j}{0} \enspace = \enspace 
\begin{cases}
\States , & \mbox{if } i > j \\
\emptyset , & \mbox{o.w.}
\end{cases}$
\end{itemize}
\end{lem}

\proof
(a) The binary representation of a number is unique. Hence, $i=j$ iff $\repr{i}{0} = \repr{j}{0}$ iff for all $s \in \States$:
$s \in \repr{i}{0}$ $\Leftrightarrow$ $s \in \repr{j}{0}$. The rest follows from the fact that $\Mubox{\mathit{test}}(I \leftrightarrow J)$ 
either holds in all states or in none of $\States$. 

(b) Similarly, we have $i < j$ iff there is a bit that is unset in $\repr{i}{0}$ but set in $\repr{j}{0}$, and $\repr{i}{0}$ and $\repr{j}{0}$ 
agree on all higher bits. Again, each disjunct of the form $\Mubox{\mathit{test}}\ldots$ is satisfied by either all or no states, and so is
the entire disjunction.

(c) Follows directly from (b). \qed

We will call an \HFL formula $\psi$ of some type $\sigma_1 \to \ldots \to \sigma_m \to \Pr$ \emph{2-valued} iff for all 
$x_1 \in \semb{\sigma_1},\ldots,x_m \in \semb{\sigma_m}$ we have $\semb{\psi}\; x_1\ldots x_m = \States$ or 
$\semb{\psi}\; x_1\ldots x_m = \emptyset$. For example, $\mathit{eq}^0$, $\mathit{lt}^0$, and $\mathit{gt}^0$ are 2-valued.
Such functions will be used to model predicates, i.e.\ functions whose return value should be either \emph{true} or \emph{false}.

Before we can extend the incrementation and decrementation functions to types $\tau_k$ for some $k > 0$ we need to define some
auxiliary functions and macros.

For \HFL formulas $\beta,\psi_1,\psi_2$ of type $\Pr$ let
\begin{displaymath}
\mathtt{if} ~ \beta ~ \mathtt{then} ~ \psi_1 ~ \mathtt{else} ~ \psi_2 \enspace := \enspace
(\beta \wedge \psi_1) \vee (\neg \beta \wedge \psi_2)
\end{displaymath} 
Note that if $\semb{\beta}$ is either $\States$ or $\emptyset$ then we have
\begin{displaymath}
\semb{\mathtt{if} ~ \beta ~ \mathtt{then} ~ \psi_1 ~ \mathtt{else} ~ \psi_2} \enspace = \enspace 
\begin{cases}
\semb{\psi_1} , & \mbox{ if } \semb{\beta} = \States \\
\semb{\psi_2} , & \mbox{ if } \semb{\beta} = \emptyset
\end{cases}
\end{displaymath}
For any $k \in \Nat$ we can easily define formulas $\mathit{min}^k$ and $\mathit{max}^k$ that encode the minimal and maximal element
in the range of $0,\ldots,F_k(p(n))-1$.
\begin{displaymath}
\begin{array}{rclrcl}
\mathit{min}^0 & := & \false \hspace{2cm} & 
\mathit{max}^0 & := & \true \\[4mm]
\mathit{min}^{k+1} & := & \lambda X.\mathit{min}^0 \hspace{2cm} &
\mathit{max}^{k+1} & := & \lambda X.\mathit{max}^0
\end{array}
\end{displaymath}
We will define by simultaneous induction on $k$ the following formulas.
\begin{itemize}
\item $\mathit{exists}^k: (\tau_k \to \Pr) \to \Pr$ \\
      It takes a predicate $P$ on the number representations  $\repr{0}{k},\ldots,\repr{F_k(p(n))-1)}{k}$ and decides whether or not there is an $i$ 
      s.t.\ $P\; \repr{i}{k}$ holds. If $P$ is 2-valued then so is $\mathit{exists}^k$. It is defined as
\begin{displaymath}
\mathit{exists}^k \enspace := \enspace \lambda P.\Big(\big(\mu Z. \lambda X. (P \; X) \vee Z \; (inc^k \; X)\big) \; \mathit{min}^k\Big)\\
\end{displaymath}

\item $\mathit{forall}^k: (\tau_k \to \Pr) \to \Pr$ \\
      Similarly, this function checks whether $P\; \repr{i}{k}$ holds for all such $i$.
\begin{displaymath}
\mathit{forall}^k \enspace := \lambda P.\neg \big((\mathit{exists}^k)\; (\neg P)\big)
\end{displaymath}

\item $\mathit{eq}^k : \tau_k \to \tau_k \to \Pr$ \\
      This is a 2-valued function which decides whether two given representations from $\repr{0}{k},\ldots,\repr{F_k(p(n))-1}{k}$ encode the same number.
      Note that for $k=0$ this has already been defined above.
\begin{displaymath}
\mathit{eq}^k \enspace := \enspace \lambda I. \lambda J. \mathit{forall}^{k-1}\; \big(\lambda X.\mathit{eq}^0 \; (I \; X)\; (J \; X)\big)
\end{displaymath}

\item $\mathit{lt}^k : \tau_k \to \tau_k \to \Pr$ \\
      This 2-valued function decides for two number representations whether the less-than-relationship holds between the two
      encoded numbers. Again, the case of $k=0$ has been dealt with above.
\begin{align*}
\hspace*{1cm}\mathit{lt}^k \enspace := \enspace \lambda I. \lambda J.
  \mathit{exists}^{k-1}\; \bigg(\lambda X.&\big(\mathit{lt}^0 \; (I \; X)\; (J \; X)\big) \enspace \wedge \\
&\mathit{forall}^{k-1}\; \Big(\lambda Y. (\mathit{gt}^{k-1} \; X \; Y) \to \big(\mathit{eq}^0 \; (I \; X)\; (J \; Y)\big)\Big)\bigg)
\end{align*}

\item $\mathit{gt}^k : \tau_k \to \tau_k \to \Pr$ \\
      Using the last one we can easily decide for two number representations whether the 
      greater-than-relationship holds between the two encoded numbers.
\begin{displaymath}
\mathit{gt}^k \enspace := \enspace \lambda I. \lambda J.(\mathit{lt}^k\; J\; I)
\end{displaymath}

\item $\mathit{inc}^k : \tau_k \to \tau_k$ \\
      This function models increment in the range of $0,\ldots,F_k(p(n))-1$ for $k > 0$.
\begin{align*}
\mathit{inc}^k \enspace := \enspace \lambda I. \lambda X.
&\mathtt{if} \enspace \mathit{exists}^{k-1}\; \big(\lambda Y. (\mathit{lt}^{k-1} \; Y \; X) \wedge \neg(\mathit{eq}^0 \; (I \; Y) \; \mathit{max}^0)\big) \\
&\mathtt{then} \enspace I \; X \enspace \mathtt{else} \enspace \mathit{inc}^0 \; (I \; X)
\end{align*}
Incrementation is done in the same way as with the binary represenation in the case of $k=0$ above: $\mathit{inc}^k$ applied to $\repr{i}{k}$
yields the function that agrees with $\repr{i}{k}$ on all arguments for which there is a smaller one whose value is not maximal, i.e.\ still
less than $F_0(p(n))-1$. If all smaller arguments including itself have already reached the maximal value then they are reset to the
minimal, i.e.\ $\repr{0}{0}$. Note that $\mathit{inc}^0$ also models the increment modulo $F_0(p(n))$. 

\item $\mathit{dec}^k : \tau_k \to \tau_k$ \\
      Similarly, this models decrement in the range of $0,\ldots,F_k(p(n))-1$ for $k > 0$.
\begin{align*}
\mathit{dec}^k \enspace := \enspace \lambda I. \lambda X.
&\mathtt{if} \enspace \mathit{exists}^{k-1}\; \big(\lambda Y. (\mathit{lt}^{k-1} \; Y \; X) \wedge \neg(\mathit{eq}^0 \; (I \; Y) \; \mathit{min}^0)\big) \\
&\mathtt{then} \enspace I \; X \enspace \mathtt{else} \enspace \mathit{dec}^0 \; (I \; X)
\end{align*}
\end{itemize}
These definitions are well-defined. For $k > 0$, $\mathit{inc}^k$ and $\mathit{dec}^k$ need $\mathit{lt}^{k-1}$,
and $\mathit{exists}^{k-1}$. The latter only needs $\mathit{inc}^{k-1}$. The former needs $\mathit{exists}^{k-2}$ and $\mathit{forall}^{k-2}$, etc.

\begin{rem} For all $k \in \Nat$ we have: 
\begin{center}
\begin{tabular}{lclclcl}
$\order{\mathit{exists}^k}$ & $=$ & $\order{\mathit{forall}^k}$ & $=$ &                         &     & $k+2$ \\
$\order{\mathit{eq}^k}$     & $=$ & $\order{\mathit{lt}^k}$     & $=$ & $\order{\mathit{gt}^k}$ & $=$ & $k+1$ \\
$\order{\mathit{inc}^k}$    & $=$ & $\order{\mathit{dec}^k}$    & $=$ &                         &     & $k+1$ \\
$\mar{\mathit{exists}^k}$   & $=$ & $\mar{\mathit{forall}^k}$   & $=$ &                         &     & $1$ \\
$\mar{\mathit{eq}^k}$       & $=$ & $\mar{\mathit{lt}^k}$       & $=$ & $\mar{\mathit{gt}^k}$   & $=$ & $2$ \\
$\mar{\mathit{inc}^k}$      & $=$ & $\mar{\mathit{dec}^k}$      & $=$ &                         &     & $2$
\end{tabular}
\end{center}
\end{rem}

The following lemmas provide exact specifications for the functions above and prove that their implementations comply to these specifications.
They are all proved by simultaneous induction on $k$.

\begin{lem}
\label{lem:existsforallcorrect}
For any function $\psi$ of type $\tau_k \to \Pr$ s.t.\ $\semb{\psi}$ is two-valued we have:
\begin{align*}
a) \quad \semb{\mathit{exists}^k\; \psi} \enspace &= \enspace  
  \begin{cases}
    \States , &\mbox{ if } \exists i \in \{0,\ldots,F_k(p(n))-1\} \mbox{ s.t. } \semb{\psi}\; \repr{i}{k} = \States \\
    \emptyset , & \mbox{ o.w. }
  \end{cases} \\
b) \quad \semb{\mathit{forall}^k\; \psi} \enspace &= \enspace  
  \begin{cases}
    \States , & \mbox{ if } \forall i \in \{0,\ldots,F_k(p(n))-1\} \mbox{ s.t. } \semb{\psi}\; \repr{i}{k} = \States \\
    \emptyset , & \mbox{ o.w. }
  \end{cases}
\end{align*}
\end{lem}

\proof
We will only prove part (a), since (b) follows from it by simple propositional reasoning. Note that for any formula $\psi$ we have
\begin{displaymath}
\mathit{exists}^k\; \psi \enspace \equiv \enspace \bigvee\limits_{i \in \Nat} p\; (
\underbrace{\mathit{inc}^k (\mathit{inc}^k (\ldots (\mathit{inc}^k}_{i \mbox{\scriptsize\ times}}\; \mathit{min}^k)\ldots)))
\end{displaymath}
by fixpoint unfolding. 
The rest follows from the correctness Lemmas~\ref{lem:incdeccorrect0} and \ref{lem:incdeccorrectk} for $\mathit{inc}^k$ and the fact that $p$ 
is assumed to be 2-valued. Clearly, the disjunction over disjuncts that all are either true or false is also either true or false. \qed

\begin{lem}
\label{lem:eqltgtcorrectk}
For all $k > 0$ and $i,j \in \{0,\ldots,F_k(p(n))-1\}$ we have:
\begin{itemize}
\item[a)] $\semb{\mathit{eq}^k}\; \repr{i}{k}\; \repr{j}{k} \enspace = \enspace 
\begin{cases}
\States , & \mbox{if } i = j \\
\emptyset , & \mbox{o.w.}
\end{cases}$

\item[b)] $\semb{\mathit{lt}^k}\; \repr{i}{k}\; \repr{j}{k} \enspace = \enspace 
\begin{cases}
\States , & \mbox{if } i < j \\
\emptyset , & \mbox{o.w.}
\end{cases}$

\item[c)] $\semb{\mathit{gt}^k}\; \repr{i}{k}\; \repr{j}{k} \enspace = \enspace 
\begin{cases}
\States , & \mbox{if } i > j \\
\emptyset , & \mbox{o.w.}
\end{cases}$
\end{itemize}
\end{lem}

\proof
(a) This follows immediately from the definition of $\mathit{eq}^k$ and Lemmas~\ref{lem:existsforallcorrect} and
\ref{lem:eqltgtcorrect0}. Note that $i = j$ iff they encode the same functions according to the representation of the
previous section. Function equality, however, can easily be tested using the $\mathit{forall}$ macro to iterate through
all possible arguments and the $\mathit{eq}^0$ function to compare the corresponding values.

(b) We have $i < j$ iff $\repr{i}{k}$ is lexicographically smaller than $\repr{j}{k}$ according to the encoding of the previous section.
Now this is the case iff there is an argument $x$ s.t.\ the value of $\repr{i}{k}$ on $x$ is smaller than the value
of $\repr{j}{k}$ on $x$, and for all arguments that are greater than $x$, these two functions agree. Hence, correctness of
$\mathit{lt}^k$ follows from Lemmas~\ref{lem:existsforallcorrect}, \ref{lem:eqltgtcorrect0} and part (c) on $k-1$.

(c) Follows from (a) and (b) by propositional reasoning. \qed

\begin{lem}
\label{lem:incdeccorrectk}
For all $k > 0$ and $i \in \{0,\ldots,F_k(p(n))-1\}$ we have:
\begin{itemize}
\item[a)] $\semb{\mathit{inc}^k}\; \repr{i}{k} \enspace = \enspace \repr{i+1 \mod F_k(p(n))}{k}$,
\item[b)] $\semb{\mathit{dec}^k}\; \repr{i}{k} \enspace = \enspace \repr{i-1 \mod F_k(p(n))}{k}$.
\end{itemize}
\end{lem}

\proof
Again, we will only prove part (a) since part (b) is entirely analogous. Let $i \in \{0,\ldots,F_k(p(n))-1\}$, and
$i' := i+1 \mod F_k(p(n))$. Remember that according to the previous section, $\repr{i'}{k}$ is the lexicographically next function
after $\repr{i}{k}$. Hence, it is the function that takes an argument $x$ and returns $\repr{i}{k}\; x$ if there is a smaller argument
$y$ s.t.\ $\repr{i}{k}\; y$ is not the maximal value. If there is no such smaller $y$ then it returns the value of $\repr{i}{k}$ on 
$x$ increased by one. This makes use of the fact that $\mathit{inc}^0$ increases modulo $F_0(p(n))$. Hence,
on all lower-valued arguments the function values are reset to $\repr{0}{0}$ again. Therefore, correctness follows from 
Lemmas~\ref{lem:incdeccorrect0}, \ref{lem:existsforallcorrect}, \ref{lem:eqltgtcorrect0}, and \ref{lem:eqltgtcorrectk}.
\qed

This provides all the necessary tools to model the behaviour of the space-bounded alternating Turing Machine $\mathcal{M}$.
In particular, $\mathit{inc}^k$ and $\mathit{dec}^k$ can be used to model the movements of the tape head on a tape of
size $F_k(p(n))$.

Remember that a configuration of $\mathcal{M}$ in the computation on $w$ is a triple $(q,h,t)$ where $q \in Q$,
$h \in \{0,\ldots,F_k(p(n))-1\}$, and $t: \{0,\ldots,F_k(p(n))-1\} \to \Gamma$. We will use the \hfl{} type $\tau_{k}$ to
model head positions $h$, and the type $\tau_{k+1}$ to model tape contents $t$. The state component of a configuration will be 
encoded in the formula. The two alphabet symbols $\true$ and $\false$ will be interpreted by the whole, resp.\ empty set
of states, i.e.\ like the propositions $\true$ and $\false$.

First of all we need to define formulas that encode the starting configuration. Formula $\mathit{head}^k_0$ encodes 
position $0$ on a tape of length $F_k(p(n))$. This is simply $\mathit{head}^k_0 := \mathit{min}^k$.

\begin{rem}
$\order{\mathit{head}^k_0} = k$, $\mar{\mathit{head}^k_0} = 1$ if $k \ge 1$ and $0$ otherwise.
\end{rem}

In order to encode the tape content of the starting configuration we need yet another auxiliary macro. 
Let $m \in \Nat$ and $j_1,\ldots,j_m$ be \hfl{} formulas of type $\tau_{k}$, and $\psi,\psi_1,\ldots,\psi_m$ be \hfl{} formulas
of type $\tau_0$. We write 
\begin{displaymath}
\mathtt{case}^k\ j_1: \psi_1,\ldots, j_m: \psi_m\ \mathtt{else}\ \psi
\end{displaymath}
to abbreviate
\begin{displaymath}
\lambda I.\bigg(\big(\bigvee\limits_{h=1}^m (\mathit{eq}^{k}\; I\; j_h) \wedge \psi_h\big) \vee 
\big(\psi \wedge \bigwedge\limits_{h=1}^m \neg(\mathit{eq}^k\; I\; j_h) \vee \neg\psi_h\big)\bigg)
\end{displaymath}
Lemma~\ref{lem:eqltgtcorrectk} immediately gives us the following. Given formulas
$j_1,\ldots,j_m$ of type $\tau_k$ that represent pairwise different numbers from $\{0,\ldots,F_k(p(n))-1\}$, and formulas 
$\psi,\psi_1,\ldots,\psi_m$, as well as a number $i \in \{0,\ldots,F_k(p(n))-1\}$, we have
\begin{displaymath}
\semb{\big(\mathtt{case}^k\ j_1: \psi_1,\ldots, j_m: \psi_m\ \mathtt{else}\ \psi\big)}\; \repr{i}{k} \enspace = \enspace
\begin{cases}
\semb{\psi_h} , & \mbox{if } \repr{i}{k} = \semb{j_h} \\
\semb{\psi} , & \mbox{o.w.}
\end{cases}
\end{displaymath}
In order to define the tape content of the starting configuration of length $F_k(p(n))$ let $w = a_0\ldots a_{n-1}$ with 
$a_i \in \{\true,\false\}$. We will use the \texttt{case}-construct to define the
initial tape content by case distinction. In order to do so, we need to explicitly address the first $n$ tape cells via
a formula $\chi^k_i$ s.t.\ $\semb{\chi^k_i} = \repr{i}{k}$ for all $i,k$. This can be done recursively using the auxiliary 
formulas $\mathit{bit}_i$ from above.
\begin{displaymath}
\chi_i^0 \enspace := \enspace \bigvee\limits_{j = 0}^{p(n)-1}
\begin{cases}
 \mathit{bit}_j \ , & \mbox{the } j\mbox{-th bit of } i \mbox{ is set} \\
 \neg\mathit{bit}_j \ , & \mbox{the } j\mbox{-th bit of } i \mbox{ is unset}
\end{cases}
\end{displaymath}
Note that here we need to represent a number in the range of $0,\ldots,F_0(p(n))-1$ by the union over all its bit values,
hence a disjunction rather than a conjunction which might seem more intuitive.

For $k > 0$ also recall that we have assumed $p(n) > n$ for all $n \in \Nat$, in particular $n \le 2^{p(n)}$. 
This ensures an easy encoding of the small numbers 
$0,\ldots,n-1$ as functions of type $\tau_{k+1}$. Function $\repr{i}{k}$, for $i \in \{0,\ldots,n-1\}$, maps $\repr{0}{k-1}$ to $\repr{i}{0}$ and all 
other arguments to $\repr{0}{0}$ -- cf.\ Fig.~\ref{fig:encode}.
Hence, for $k > 0$, let 
\begin{displaymath}
\chi_i^k \enspace := \enspace \mathtt{case}\ \mathit{min}^{k-1}: \chi^0_i\ \mathtt{else}\ \mathit{min}^0
\end{displaymath}
This allows us to represent the starting configuration of $\mathcal{M}$ on $w$ as a simple case distinction. 
\begin{displaymath}
\mathit{tape}_0^k \enspace := \enspace \mathtt{case}^k\ \chi_0^k: a_0, \ldots, \chi_{n-1}^k: a_{n-1}\ \mathtt{else}\ \false
\end{displaymath}
Here we utilise the fact that we encode the alphabet symbols $\true$ and $\false$ using the propositions $\true$ and $\false$
and the blank tape by a sequence of the symbol $\false$.

\begin{rem}
$\order{\mathit{tape}_0^k} = k+1$, $\mar{\mathit{tape}_0^k} = 2$.
\end{rem}

Next we need formulas that encode the manipulation of configurations. In particular, we will have to model the head movement, 
and define formulas for reading and updating the symbol at a certain tape position. 
Remember that in an $F_k(p(n))$-space bounded configuration, the head position
can be encoded using type $\tau_{k}$, and the tape content can be encoded using type $\tau_{k} \to \Pr = \tau_{k+1}$. We need to define the
following functions for any $a \in \Gamma$.
\begin{itemize}
\item $\mathit{read}_a^k : \tau_{k+1} \to \tau_k \to \Pr$ \\
      Applied to an encoded tape content and head position it tests whether or not the symbol under the head on that tape is $a$.
      It is also a 2-valued predicate. Remember that there only are the two symbols $\true$ and $\false$ with corresponding encoding.
\begin{align*}
\mathit{read}_\true^k \enspace &:= \enspace \lambda T.\lambda H.(T\; H) \\
\mathit{read}_\false^k \enspace &:= \enspace \lambda T.\lambda H.\neg(T\; H)
\end{align*}

\item $\mathit{write}_a^k : \tau_{k+1} \to \tau_{k} \to \tau_{k+1}$ \\ 
      Given an encoded tape content $t$ and a head position $h$ it returns the tape content that contains $a$ at position $h$ and
      complies with $t$ on all other positions.
\begin{displaymath}
\mathit{write}_a^k \enspace := \enspace \lambda T.\lambda H.\lambda H'.\mathtt{if}\ (\mathit{eq}^k\; H\; H')\ 
\mathtt{then}\ a\ \mathtt{else}\ (T\; H')
\end{displaymath}
\end{itemize}

\begin{rem}
$\order{\mathit{read}^k_a} = k+2$, $\order{\mathit{write}_a^k} = k+2$, $\mar{\mathit{read}_a^k} = 2$, $\mar{\mathit{write}^k_a} = 3$.
\end{rem}

In the following we write $\repr{t}{k}$ for the encoding of the tape $t$ of length $F_k(p(n))$ as a function of type $\tau_k \to \Pr$.
Equally, the head position $h$ in a configuration is encoded by $\repr{h}{k}$. We also write $t[h := a]$ for the update of $t$ with $a$ 
at position $h$. The next two lemmas show that the above functions are correct. Their proofs are straight-forward. The latter relies
on the correctness of the \texttt{if-then-else}-construct.

\begin{lem}
\label{lem:readcorrect}
For all $k \in \Nat$, all $x \in \Gamma$, all tape contents $t$, and all head positions $h$ we have
\begin{displaymath}
\semb{\mathit{read}_x^k}\; \repr{t}{k}\; \repr{h}{k} \enspace = \enspace
\begin{cases}
\States , & \mbox{if the symbol in } t \mbox{ at position } h \mbox{ is } x\\
\emptyset , & \mbox{o.w.}
\end{cases}
\end{displaymath}
\end{lem}

\begin{lem}
\label{lem:writecorrect}
For all $k \in \Nat$, all $x \in \Gamma$, all tape contents $t,t'$, and all positions $h$ we have:
$\semb{\mathit{write}_x^k}\; \repr{t}{k}\; \repr{h}{k} = \repr{t'}{k}$ iff $t' = t[h := x]$.
\end{lem}

The movement of the tape head is easily modeled using three functions $\mathit{move}^k_d : \tau_k \to \tau_k$ for
$d \in \{-1,0,+1\}$.
\begin{displaymath}
\mathit{move}^k_{-1} \enspace := \enspace \mathit{dec}^k \enspace, \quad \mathit{move}^k_0 \enspace := \enspace \lambda H.H
\enspace , \quad \mathit{move}^k_{+1} \enspace := \enspace \mathit{inc}^k
\end{displaymath}
Finally, we use the characterisation of acceptance in an alternating Turing Machine as a reachability game to construct
the formula $\Phi^k_{\mathcal{M},w}$. Let $Q = \{q_0,\ldots,q_m,q_{\mathit{acc}},q_{\mathit{rej}}\}$. We will simultaneously
define for each state $q \in Q$ an eponymous function $q: \tau_{k+1} \to \tau_{k} \to \Pr$ that -- given a tape content
$t$ and a head position $h$ -- signals as a 2-valued predicate whether or not $\mathcal{M}$ accepts starting in the configuration $(q,h,t)$.
Let, for all $q \in Q$,
\begin{displaymath}
\Psi^k_{\mathcal{M},q} \enspace := \enspace \mu q.\left(
\begin{array}{lcl}
q_0 & . & \lambda T.\lambda H.\Psi_0 \\
&\, \vdots & \\
q_m & . & \lambda T.\lambda H.\Psi_m \\
q_{\mathit{acc}} & . & \lambda T.\lambda H.\true \\
q_{\mathit{rej}} & . & \lambda T.\lambda H.\false
\end{array}\right)
\end{displaymath}
where for all $i = 0, \ldots,m$:
\begin{displaymath}
\Psi_i \enspace := \enspace \bigvee\limits_{a \in \Gamma} (\mathit{read}^k_a\; T\; H) \wedge  
\begin{cases}
\bigvee\limits_{(q',b,d) \in \delta(q_i,a)}\hspace*{-6mm} 
  q'\; (\mathit{write}_b^k\; T\; H)\; (\mathit{move}_d^k\; H) &, \mbox{ if } q \in Q_\exists \\
\bigwedge\limits_{(q',b,d) \in \delta(q_i,a)}\hspace*{-6mm} 
  q'\; (\mathit{write}_b^k\; T\; H)\; (\mathit{move}_d^k\; H) &, \mbox{ if } q \in Q_\forall \\
\end{cases}
\end{displaymath}
Then define $\Phi^k_{\mathcal{M},w} := \Psi_{\mathcal{M},q_0}^k\; \mathit{tape}^k_0\; \mathit{head}^k_0$.

The following result about the order-restricted fragment into which $\Phi_{\mathcal{M},w}^k$ falls is easily obtained
by collecting all the preceding remarks about the orders and maximal arities of all its subformulas. Note that those of highest 
type-order are $\mathit{read}^k_a$, $\mathit{write}^k_a$, and $q$ for each $q \in Q$. All of them have order $k+2$.

\begin{lem}
\label{lem:typeofphik}
For all $k \in \Nat$: $\Phi_{\mathcal{M},w}^k \in \hfl{k+2,3}$.
\end{lem}

\begin{thm}
\label{thm:lowerboundcorrect}
For all $k \ge 1$, all $w \in \Gamma^*$ and all $F_k(p(n))$-space bounded alternating Turing Machines $\mathcal{M}$ we have:
\begin{displaymath}
\semb{\Phi^k_{\mathcal{M},w}}^{\Transsys_{\mathcal{M},w}} \enspace = \enspace
\begin{cases}
\States , & \mbox{if } w \in L(\mathcal{M}) \\
\emptyset, & \mbox{o.w.}
\end{cases}
\end{displaymath}
\end{thm}

\proof
Let $\mathcal{M} = (Q,\Sigma,\Gamma,q_0,\delta,q_{\mathit{acc}},q_{\mathit{rej}})$. Suppose $w \in L(\mathcal{M})$. Then there is 
an accepting run of $\mathcal{M}$ on $w$. Remember that $\mathcal{M}$ is alternating. Hence, this run can be represented as a tree 
$T$ with starting configuration $(q_0,0,w\false\ldots\false)$ as the root, s.t.\ 
\begin{itemize}
\item every existential configuration has exactly one successor in the tree,
\item for every universal configuration the set of its successors in the tree forms the set of all its successor configurations,
\item all leaves are accepting configurations.
\end{itemize}
We now show $\semb{\Phi^k_{\mathcal{M},w}}^{\Transsys_{\mathcal{M},w}} = \States$ by induction on the height $h(T)$ of $T$. Since
$\Phi^k_{\mathcal{M},w}$ is a simultaneously defined fixpoint function applied to two arguments we need a stronger inductive
hypothesis. We will show that for all $q \in Q$ and all $t: \tau_k \to \Pr$, all $h: \tau_k$ encoding a tape content and a head position: 
$\semb{\Psi_{\mathcal{M},q}^k}\; \repr{t}{k}\; \repr{h}{k} = \States$ if $\mathcal{M}$ accepts starting in the configuration given by $(q,h,t)$.

The base case is $h(T) = 1$ which means that the root is an accepting configuration. Hence, $q = q_{\mathit{acc}}$, and the
claim is easily seen to be true by two applications of $\beta$-reduction.

If $h(T) > 1$ then we need to distinguish two cases. First, assume that $q \in Q_\exists$. Then there is exactly one successor
configuration $(q',t',h')$ in $T$ which results from $(q,t,h)$ by one Turing Machine step according to $\delta$. Clearly, $\mathcal{M}$
accepts starting in $(q',t',h')$ and, by hypothesis, we have $\semb{\Psi_{\mathcal{M},q'}^k}\; \repr{t'}{k}\; \repr{h'}{k} = \States$. One unfolding of the
fixpoint formula together with Lemmas~\ref{lem:incdeccorrect0}, \ref{lem:incdeccorrectk}--\ref{lem:writecorrect} show that we
also have $\semb{\Psi_{\mathcal{M},q}^k}\; \repr{t}{k}\; \repr{h}{k} = \States$.

The case of $q \in Q_\forall$ is similar. Here, there are possibly several accepting subtrees of $T$. But the
hypothesis applies to all of them and intersection over $\States$ several times is still $\States$.

This shows completeness. Soundness can be proved along the same lines because of determinacy. Note that if $w \not\in L(\mathcal{M})$
then this is witnessed by a computation tree in which every universal configuration has only one successor, every existential
one retains all of its successors, and all leaves are rejecting. \qed

\subsection{Lower Bounds on the Model Checking Complexity}

\begin{thm}
\label{thm:kexptimehard}
For all $k \ge 2$ and all $m \ge 3$ the model checking problem for \hfl{k,m} is $\EXPTIME{k}$-hard when $|\Prop| \ge 0$, $|\Act| \ge 2$. 
\end{thm}

\proof
Let $k \ge 2$. According to Thm.~\ref{thm:alternation} there is a \AEXPSPACE{(k-1)} machine $\mathcal{M}$ s.t.\ $L(\mathcal{M})$ is
$\EXPTIME{k}$-hard \cite{Chandra:1981:A}. Using padding we can assume the space required by $\mathcal{M}$ on an input word of length $n$ 
to be bounded by $F_{k-2}(p(n))$ for some polynomial $p(n) > n$. Thm.~\ref{thm:lowerboundcorrect} yields a reduction from 
$w \in \Gamma^*$ to labeled transition systems $\Transsys_{\mathcal{M},w}$ and a formula $\Phi_{\mathcal{M},w}^{k-2}$ s.t.\ 
$w \in L(\mathcal{M})$ iff $\Transsys_{\mathcal{M},w}, s \models \Phi^{k-2}_{\mathcal{M},w}$ for any state $s$. 

According to Lemma~\ref{lem:typeofphik} we have $\Phi_{\mathcal{M},w}^{k-2} \in \hfl{k,3}$. Furthermore, $|\Transsys_{\mathcal{M},w}|$ 
is clearly polynomial in $n$. The size of $\Phi^{k-2}_{\mathcal{M},w}$ is also
polynomial in $n$, but this formula is only
an abbreviation using the simultaneous fixpoint definition in $\Phi^{k}_{\mathcal{M},q_0}$ and we need to consider the Fisher-Ladner closure of its unabbreviated counterpart. But remember the definition
of $\Phi^{k-2}_{\mathcal{M},w}$ as $\Psi_{\mathcal{M},q_0}^k\; \mathit{tape}^k_0\; \mathit{head}^k_0$. Unfolding only affects the
subformula $\Psi_{\mathcal{M},q_0}^k$ whose size is independent of $n$. Hence, $|\Phi^{k-2}_{\mathcal{M},w}|$ is also polynomial in $n$.
\qed

For the fragment \hfl{1} a similar result follows from the known \ExpTime lower bound for \FLC \cite{lange:3mcflc:05} and the
embedding of \FLC into \hfl{1,1} \cite{viswanathan:hfl}.

\begin{prop}[\cite{lange:3mcflc:05,viswanathan:hfl}]
There is an \hfl{1,1} formula over a singleton $\Prop$ and an $\Act$ of size $2$ whose set of models is \ExpTime-hard. 
\end{prop}

The condition $|\Act| \ge 2$ results from the fact that the reduction to \FLC model checking is from the pushdown game
problem. The number of different modal accessibility relations is $p+1$ where $p$ is the size of the alphabet in the
pushdown games. A close inspection of the \ExpTime lower bound proof for this problem \cite{Walukiewicz01} shows that $p=1$ is sufficient.

It has already been observed that model checking \HFL on fixed and very small transition systems is non-elementary
\cite{langesomla-mfcs05}. We repeat this observation here since it follows from the construction above in a very neat 
way. Remember that $\log^* n = i$ iff the $i$-fold iteration of the function $\lambda m.\lceil \log m\rceil$ starting 
in $n$ yields $1$. 
 
\begin{thm}
\label{thm:exprcomplnonelem}
The model checking problem for \HFL on the fixed transition system of size $1$, no transitions and no 
labels is non-elementary when maximal type arities are at least $2$.
\end{thm}

\proof
Note that Thm.~\ref{thm:lowerboundcorrect} uses $p(n)$ many states to encode $F_k(p(n))$ many numbers for any $k \ge 1$. But
$F_k(p(n)) = 2^{p(n)\cdot F_{k-1}(p(n))}$, thus $F_{k+1}(\log p(n)) \ge F_k(p(n))$. This means that the
reduction in Thm.~\ref{thm:lowerboundcorrect} also works with $\log p(n)$ many states, but yields a formula in \hfl{k+1} 
rather than \hfl{k}. Iterating this shows that one state suffices for the reduction, but the result is only in
\hfl{k + \log^* p(n)}.

Finally, note that by the construction above, this single state $0$ does not have any $\mathit{lower}$-transitions. The 
transition $\Transition{0}{\mathit{test}}{0}$ is redundant because we have $0 \models \Mubox{\mathit{test}}\psi$ iff 
$0 \models \psi$ for any formula $\psi$.
\qed

There is an apparent intuitive mismatch between this and Cor.~\ref{cor:exprcomplupper} which both make a statement about
the expression complexity of \HFL on the smallest possible transition system. For every fixed $k,m$, this is in \ExpTime.
However, when $k$ is unbounded it becomes non-elementary. Even though this gap is huge in terms of complexity classes
it is just tiny in terms of the \HFL types that are necessary to achieve a non-elementary complexity: the type levels 
only have to be increased by $\log^* p(n)$. Note that $\log^* m \le 6$ for any natural number $m$ that is
representable using electron spins as bits when the entire observable universe was densely packed with electrons. 
The cause for the apparent intuitive mismatch is simply an underestimation of the exponential time hierarchy. Equally,
a tower of height $6$ is sufficient to exceed the numbers representable using the electron spins in this way.
 
The above two theorems raise the question after a lower bound for the data complexity of \HFL. In the following we will 
modify the reduction to yield a formula $\Phi^k_{\mathcal{M}}$ that only depends on the alternating
Turing Machine $\mathcal{M}$ rather than both the machine and the input word. Remember that, according to 
Thm.~\ref{thm:alternation}, there is -- for any $k \ge 1$ -- such a machine with a word problem that is
\EXPTIME{k}-hard.

The idea for the modification is simple. It is only the subformula $\mathit{tape}^k_0$ that depends on the input word.
First, let 
\begin{displaymath}
\mathit{tape}^k_{\mathit{empty}} \enspace := \enspace \lambda X.\false
\end{displaymath}
model the tape that contains the blank symbol $\Box$ only. 

\begin{figure}[t]
\framebox[\textwidth]{
\parbox{\textwidth}{
\begin{center}
\scalebox{0.6}{\includegraphics{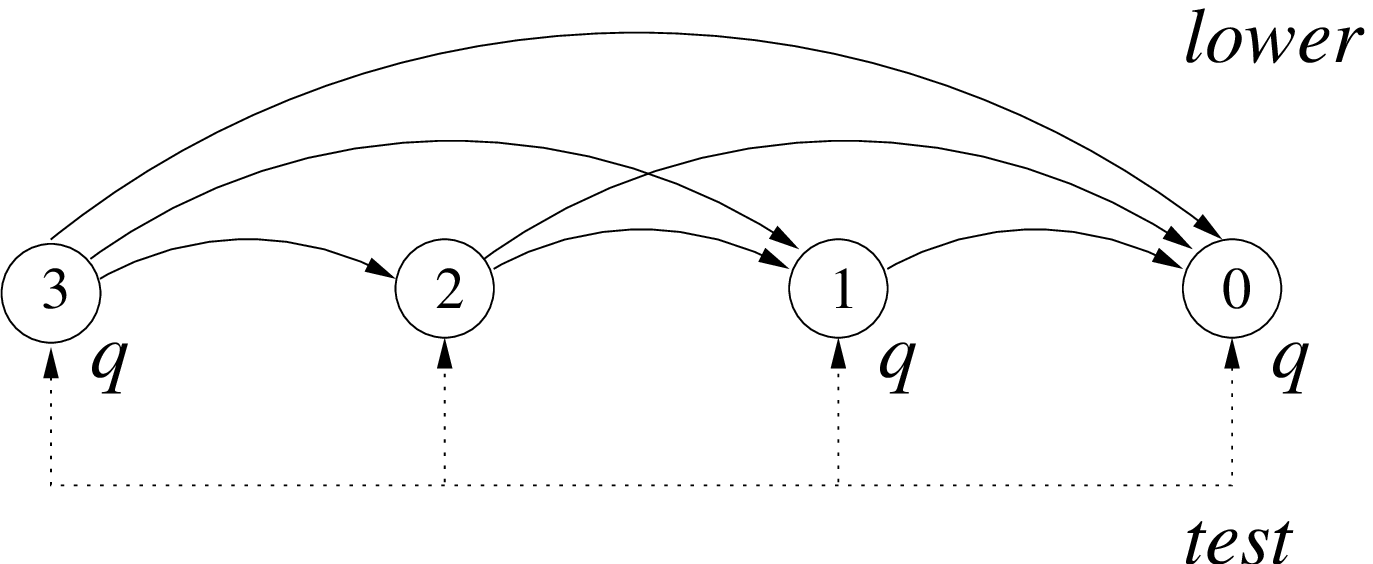}}
\end{center}
}}
\caption{The transition system $\Transsys'_{\mathcal{M},w}$ for $p(n) = 4$ and $w = \true\,\true\,\false\,\true$.}
\label{fig:transsys2}
\end{figure}

Note that $\Transsys_{\mathcal{M},w}$ has $p(n) > n$ many states. Hence, we can use these states together with a
single proposition $q$ to model the input word $w = a_0,\ldots,a_{n-1} \in \{\true,\false\}^*$.
\begin{displaymath}
L(i) \enspace = \enspace 
\begin{cases}
\{q\}\ , & \mbox{if } i \le n \mbox{ and } a_i = \true \\
\emptyset \ , & \mbox{o.w. }
\end{cases}
\end{displaymath}
Let $\Transsys'_{\mathcal{M},w}$ be the result of this. Note that it differs from $\Transsys_{\mathcal{M},w}$ only 
through the additional labels on the states. An example with $p(n) = 4$ and $w = \true\,\true\,\false\,\true$ is 
shown in Fig.~\ref{fig:transsys2}. For better readability we depict the relation $\Transition{}{\mathit{test}}{}$ only
schematically.

All we need now is a formula that traverses through these states and uses the information obtained from each label to
generate the original encoding $\mathit{tape}^k_0$ of the real input tape. This is done by the function 
$\mathit{build}^k: \tau_{k+1} \to \tau_k \to \Pr \to \Pr \to \tau_{k+1}$, defined as
\begin{align*}
\mathit{build}^k \enspace := \enspace \mu Z.&\lambda T.\lambda H.\lambda C.\lambda Y. \\
&\mathtt{if}\ \Mubox{test}(\chi \leftrightarrow \mathit{bit}_{p(n)}) \ \mathtt{then}\ T \\
&\mathtt{else\ if}\ \Mubox{test}(C \to q) \\
&\hspace*{9mm}\mathtt{then} \ Z\; (\mathit{write}^k_{\true}\; T\; H)\; (\mathit{inc}^{k-1}\; H)\; (\Mudiam{lower}C \wedge \Mubox{\mathit{lower}}Y)\; (C \vee Y) \\
&\hspace*{9mm}\mathtt{else} \ Z\; (\mathit{write}^k_{\false}\; T\; H)\; (\mathit{inc}^{k-1}\; H)\; (\Mudiam{lower}C \wedge \Mubox{\mathit{lower}}Y)\; (C \vee Y)
\end{align*}
The parameters $T$ and $H$ contain the current tape content and the position at which the next symbol is going to be written.
From this perspective it is not surprising that $\mathit{write}^k$ and $\mathit{inc}^k$ are applied to them before a recursive
call of $Z$. The parameters $C$ and $Y$ are used to identify the next state in $\Transsys_{\mathcal{M},w}$ which is 
checked for the label $q$. Remember the recursive definition of the formulas $\mathit{bit}_i$ which is exactly what is reproduced
here. Note that $\order{\mathit{build}^k} = k+2$ and $\mar{\mathit{build}^k} = 5$. Finally, let
\begin{displaymath}
\mathit{tape}^k_{\mathit{built}} \enspace := \enspace \mathit{build}^k\; \mathit{tape}^k_{\mathit{empty}}\; \mathit{min}^k\; \mathit{bit}_{0}\; \false
\end{displaymath}

\begin{lem}
\label{lem:build}
For all $\Transsys'_{\mathcal{M},w}$ which, in addtion to $\Transsys_{\mathcal{M},w}$ carry the input word $w$ through labels
as defined above we have $\semb{\mathit{tape}^k_{\mathit{built}}}^{\Transsys'_{\mathcal{M},w}} = \semb{\mathit{tape}^k_0}^{\Transsys_{\mathcal{M},w}}$.
\end{lem}

\proof
Assume that $\repr{t}{k}$ encodes a tape content $t$ and $\repr{h}{k}$ a head position on this tape. Then we have for all $i \in \{0,\ldots,p(n)-1\}$:
\begin{align*}
\semb{\mathit{build}^k}\; &\repr{t}{k}\; \repr{h}{k}\; \semb{\mathit{bit}_i}\; \semb{\bigvee\nolimits_{j=0}^{i-1} \mathit{bit}_j} \enspace = \\
&\begin{cases}
\repr{t}{k} \ , & \mbox{if } i=p(n) \\
\semb{\mathit{build}^k}\; \repr{t[h:=a]}{k}\; \repr{h+1}{k}\; \semb{\mathit{bit}_{i+1}}\; \semb{\bigvee_{j=0}^{i} \mathit{bit}_j}\ , &\mbox{if } i < p(n)
\end{cases}
\end{align*}
where $a = \true$ if $L(i) = q$ and $a = \false$ if $L(i) = \emptyset$. Thus, when applied to the initial values encoding the blank
tape, leftmost head position, the state representing bit 0 and the empty disjunction, this least fixpoint recursion eventually yields
the tape onto which the word $w$ at hand is written. This makes use of the fact that $p(n) > n$, i.e.\ the fixpoint recursion takes 
at least one more step after reading the entire input word before it terminates.
\qed

\begin{thm}
\label{thm:datacompllower}
For all $k \ge 2$ and all $m \ge 5$ there is an \hfl{k,m} formula over a singleton $\Prop$ and an $\Act$ of size $2$ whose set of models is 
\EXPTIME{k}-hard.
\end{thm}

\proof
Let $\Phi^k_{\mathcal{M}} := \Psi^k_{\mathcal{M},q_0}\; \mathit{tape}^k_{\mathit{built}}\; \mathit{head}^k_0$. Clearly, $\Phi^k_{\mathcal{M}}$
only depends on $\mathcal{M}$ and not on its input word $w$. Furthermore, we have $\Phi^k_{\mathcal{M}} \in$ \hfl{k+2,5}. The hardness
result then follows from Lemma~\ref{lem:build} and Thm.~\ref{thm:lowerboundcorrect} along the same lines as the proof of
Thm.~\ref{thm:kexptimehard}.
\qed

It is possible to reduce the maximal arity to 4 at the cost of an extra accessibility relation in the model. If there are transitions 
$\Transition{i}{\mathit{pred}}{j}$ iff $j = i-1$ then the formulas $\mathit{bit}_i$ can be defined more simply as 
$\mathit{bit}_0 := \Mubox{\mathit{pred}}\false$ and $\mathit{bit}_{i+1} := \Mudiam{\mathit{pred}}\mathit{bit}_i$, and the parameter
$Y$ in $\mathit{build}^k$ is unnecessary.


%% file: concl.tex
\section{Conclusions}

The table in Fig.~\ref{fig:summary} shows the complexity of the model checking problem for \HFL. We distinguish the \emph{combined complexity}
(both transition system and formulas as input), the \emph{expression complexity} (model checking on a fixed transition system), 
and the \emph{data complexity} (model checking with a fixed formula). Note that lower bounds from either expression or data
complexity trivially transfer to the combined complexity while upper bounds for that trivially transfer back to both of them.
In any case, $n$ denotes the size of the transition system, $\varphi$ is the input formula, $k$ the maximal type order and
$m$ the maximal type arity of one of its subformulas, and $p := |\Prop| + |\Act|$ is the number of underlying propositions and
modal acessibility relations. Note that there are standard translations for modal logics that reduce one at the cost of 
increasing the other whilst preserving satisfiability. These could be incorporated directly into the reduction for the lower bound.

\newcommand{\msp}{\hspace*{2mm}}

\begin{figure}[t]
\begin{center}
\begin{tabular}{lc||c||c||c}
\multicolumn{2}{l||}{\rule[-2mm]{0pt}{6mm}complexity} & combined & data & expression \\ \hline\hline
\multirow{2}{*}{\raisebox{-4mm}{\HFL}}
  & \rule[-3mm]{0pt}{9mm} $\in$ & \DTIME{\tower{n\cdot |\varphi|^{O(|\varphi|)}}{k}} & \multirow{2}{*}{\raisebox{-3mm}{\EXPTIME{k}}} & \DTIME{\tower{|\varphi|^{O(|\varphi|)}}{k}} \\ \cline{2-3}\cline{5-5}
  & \rule[-3mm]{0pt}{8mm} hard & \ELTIME                          &                 & \ELTIME  \\ \hline\hline
\multirow{2}{*}{\raisebox{-4mm}{\hfl{0}}}
  & \rule[-3mm]{0pt}{9mm} $\in$ & UP$\cap$co-UP & \multirow{2}{*}{\raisebox{-3mm}{P}} & UP$\cap$co-UP \\ \cline{2-3}\cline{5-5}
  & \rule[-3mm]{0pt}{8mm} hard  & P             &                                     & P  \\ \hline\hline
\multirow{2}{*}{\raisebox{-4mm}{\hfl{1,m}}}\msp
  & \rule[-3mm]{0pt}{9mm} $\in$ 
     & \multirow{2}{*}{\msp\raisebox{-4mm}{\parbox{3cm}{\begin{center}\ExpTime\\[2mm]\scriptsize(when $p \ge 3$)\end{center}}}\msp} 
     & \multirow{2}{*}{\raisebox{-3mm}{\parbox{2.5cm}{\begin{center}\ExpTime\\[2mm]\scriptsize(when $p \ge 3$)\end{center}}}} 
     & \ExpTime \\ \cline{2-2}\cline{5-5}
  & \rule[-3mm]{0pt}{8mm} hard  &                           &                                            & P  \\ \hline\hline
\multirow{2}{*}{\raisebox{-4mm}{\hfl{k,m}, $k\ge 2$}}\msp
  & \rule[-3mm]{0pt}{9mm} $\in$ 
     & \multirow{2}{*}{\msp\raisebox{-4mm}{\parbox{3cm}{\begin{center}\EXPTIME{k}\\[2mm]\scriptsize(when $m \ge 3$, $p \ge 2$)\end{center}}}\msp} 
     & \multirow{2}{*}{\raisebox{2mm}{\parbox{3cm}{\begin{center}\EXPTIME{k}\\[1mm]\scriptsize(when $m \ge 5, p \ge 3$\\or $m \ge 4, p \ge 4$)\end{center}}}} 
     & \ExpTime \\ \cline{2-2}\cline{5-5}
  & \rule[-3mm]{0pt}{8mm} hard  &                           &                                            & P \\ \hline\hline
\end{tabular}
\end{center}

\caption{A summary of the model checking complexity results.}
\label{fig:summary}
\end{figure}

The entries stretching over two columns denote completeness results for the corresponding complexity class. The restrictions of
the form $m \ge 2$ etc.\ of course only apply to the respective lower bound.

The estimation on the time complexity of model checking general \HFL uses the fact that the maximal type order as well as the maximal 
type arity of a subformula of $\varphi$ are both bounded by $|\varphi|$.

Recall that \ELTIME does not have complete problems under polynomial time reductions. The upper bounds on the expression and combined complexity 
for general \HFL model checking are therefore as close as possible to the corresponding lower bound. 

The gaps between P and UP$\cap$co-UP simply restate open questions about the exact model checking complexity of the modal $\mu$-calculus.
The best upper bound known there in terms of complexity classes is UP$\cap$co-UP so far \cite{Jurdzinski/98}. The polynomial time lower bound 
for its expression complexity is taken from an unpublished manuscript \cite{DJN96:manuscript}. Despite a lot of effort this gap remains open up to date. 

The only question about the complexity of model checking \HFL that is left unanswered but might be feasible is the
gap in the expression complexity of \hfl{k,m} for any fixed $k,m \ge 1$. It remains to be seen whether there are fixed transition
systems $\Transsys_{k,m}$, s.t.\ for all $k,m$, the set of \hfl{k,m} formulas that are satisfied by $\Transsys_{k,m}$ is \ExpTime-hard.

Finally, the \EXPTIME{k}-completeness of \hfl{k,m}'s data complexity immediately implies a hierarchy result regarding
expressive power.

\begin{cor}
For all $k \in \Nat$ we have: \hfl{k} $\lneq$ \hfl{k+1}.
\end{cor} 

\begin{proof}
For $k=0$ this is known already because of $\hfl{0} = \mucalc{} \lneq$ FLC $\le$ \hfl{1,1} \cite{Mueller-Olm:1999:MFL,viswanathan:hfl}.
Now take any $k \ge 1$. According to Thm.~\ref{thm:kexptimehard}, there is a formula $\varphi \in \hfl{k+1,5}$ whose set of models is 
$\EXPTIME{(k+1)}$-hard. Now suppose that there is also a $\psi \in \hfl{k,m}$ for some $m \ge 1$ s.t.\ $\psi \equiv \varphi$. Note 
that $\varphi$ is fixed, and so is $\psi$. According to Thm.~\ref{cor:inkexptime}, this same set of models 
would also be included in \EXPTIME{k} which contradicts the complexity-theoretic time-hierarchy theorem of 
$\EXPTIME{k} \subsetneq \EXPTIME{(k+1)}$.
\end{proof}
